\documentclass[floatfix, reprint, superscriptaddress, aps, showpacs, nofootinbib, prb]{revtex4-2}%

\pdfoutput=1
\usepackage{graphicx}
\usepackage{dcolumn}
\usepackage{bm}                 
\usepackage{amsmath,amsbsy,amsfonts,revsymb}
\usepackage{color}
\usepackage{textcomp}
\usepackage{xcolor}
\usepackage{tikz}
\usepackage[leftcaption]{sidecap}
\usepackage{collcell}
\usepackage[utf8]{inputenc}
\def\wn/{$\,\mathrm{cm}^{-1}$}
\def\wnnospace/{cm$^{-1}$}
\def\area/{\,cm$^{-2}$}
\def\cubic/{$_\mathrm{c}$}
\def\DM/{Dzyaloshinskii-Moriya}
\def\lnpo/{LiNiPO$_4$}
\def\lcpo/{LiCoPO$_4$}
\def\lfpo/{LiFePO$_4$}
\def\lmpo/{Li$M$PO$_4$}
\def\lmnpo/{LiMnPO$_4$}

\def\sw/{spin-wave}
\def\Sw/{Spin-wave}

\def\edc/{\ensuremath{\mathbf{E}}}
\def\eac/{\ensuremath{\mathbf{E}^\omega}}
\def\hdc/{\ensuremath{\mathbf{H}}}
\def\hac/{\ensuremath{\mathbf{H}^\omega}}
\def\hac/{\ensuremath{\mathbf{H}^\omega}}
\def\kvec/{\ensuremath{{\bf k}}}
\def\Svec/{\ensuremath{{\bf S}}}
\def\ew/{\ensuremath{\mathit{E}^\omega}}
\def\bw/{\ensuremath{\mathit{H}^\omega}}
\def\tn/{\ensuremath{T_\mathrm{N}}}

\newcommand{\ewu}[1]{\ensuremath{E^\omega_{#1}}}

\newcommand{\hwu}[1]{\ensuremath{\mathit{H}^\omega_{#1}}}
\newcommand{\hparal}[1]{\textbf{H}$\,\parallel\,$\textbf{{#1}}}
\newcommand{\vect}[1]{\ensuremath{\mathbf{#1}}}
\def\ew/{\ensuremath{E^\omega}}
\def\hw/{\ensuremath{H^\omega}}
\def\bw/{\ensuremath{B^\omega}}

\def\fos/{$F_{\mathrm{os}}$} 
\def\slope/{\ensuremath{b_1}}

\newcommand{\mode}[1]{$F_{#1}$}

\makeatletter
\renewcommand{\p@subsection}{}
\renewcommand{\p@subsubsection}{}
\makeatother
\begin{document}
	\title{Refining magnetic interactions from the magnetic field dependence of spin-wave excitations in magnetoelectric \lfpo/}
	
	\author{L. Peedu}
	\affiliation{National Institute of Chemical Physics and Biophysics, Akadeemia tee 23, 12618 Tallinn, Estonia}
	
	\author{V. Kocsis}
	\affiliation{RIKEN Center for Emergent Matter Science (CEMS), Wako, Saitama 351-0198, Japan}
	
	\author{D. Szaller}
	\affiliation{Institute of Solid State Physics, TU Wien, 1040 Vienna, Austria}
	
	\author{B. Forrai}
    \affiliation{Department of Physics, Institute of Physics, Budapest University of Technology and Economics, M\H{u}egyetem rkp. 3., H-1111 Budapest, Hungary}
	
	\author{S. Bord\'acs}
    \affiliation{Department of Physics, Institute of Physics, Budapest University of Technology and Economics, M\H{u}egyetem rkp. 3., H-1111 Budapest, Hungary}
	
	\author{I. K\'ezsm\'arki}
    \affiliation{Department of Physics, Institute of Physics, Budapest University of Technology and Economics, M\H{u}egyetem rkp. 3., H-1111 Budapest, Hungary}
	\affiliation{Experimental Physics 5, Center for Electronic Correlations and Magnetism,Institute of Physics, University of Augsburg, 86159 Augsburg, Germany}
	
	\author{J. Viirok}
	\affiliation{National Institute of Chemical Physics and Biophysics, Akadeemia tee 23, 12618 Tallinn, Estonia}
	
	\author{U. Nagel}
	\affiliation{National Institute of Chemical Physics and Biophysics, Akadeemia tee 23, 12618 Tallinn, Estonia}
	
	\author{B. Bern\'ath}
    \author{D. L. Kamenskyi}
    \affiliation{High Field Magnet Laboratory (HFML-EMFL), Radboud University, Toernooiveld 7, 6525 ED Nijmegen, The Netherlands}
	
	\author{A. Miyata}
	\author{O. Portugall}
	\affiliation{Laboratoire National des Champs Magn\'etiques Intenses (LNCMI-EMFL), CNRS-UGA-UT3-INSA,143 Avenue de Rangueil, 31400 Toulouse, France}
		
	\author{Y. Tokunaga}
	\affiliation{RIKEN Center for Emergent Matter Science (CEMS), Wako, Saitama 351-0198, Japan}
	\affiliation{Department of Advanced Materials Science, University of Tokyo, Kashiwa 277-8561, Japan}
	
	\author{Y. Tokura}
	\affiliation{RIKEN Center for Emergent Matter Science (CEMS), Wako, Saitama 351-0198, Japan}
	\affiliation{Department of Applied Physics, University of Tokyo, Hongo, Tokyo 113-8656, Japan}
	
	\author{Y. Taguchi}
	\affiliation{RIKEN Center for Emergent Matter Science (CEMS), Wako, Saitama 351-0198, Japan}
	
	\author{T. R{\~o}{\~o}m}
	\affiliation{National Institute of Chemical Physics and Biophysics, Akadeemia tee 23, 12618 Tallinn, Estonia}
	
	\date{\today }

	\begin{abstract}
		
		We investigated the spin excitations of magnetoelectric \lfpo/ by THz absorption spectroscopy in magnetic fields up to 33\,T. 
		By studying their selection rules, we found not only magnetic-dipole, but also electric-dipole  active (electromagnons) and magnetoelectric resonances. 
		The magnetic field dependence of four strong low-energy modes is reproduced well by our four-sublattice spin model  for fields applied along the three orthorhombic axes.
		From the fit, we refined the exchange couplings, single-ion anisotropies, and the \DM/ interaction parameters.
		Additional spin excitations not described by the mean-field model are observed at higher frequencies.
		Some of them shows a strong shift with magnetic field, up to 4\,\wn//T, when the field is applied along the easy axis.
		Based on this field dependence, we attribute these high frequency resonances to excitation of higher spin multipoles and of two magnons, which become THz-active due to the low symmetry of the magnetically ordered state.
		
	\end{abstract}
	\maketitle
	\section{Introduction}

Recent optical studies of multiferroic materials have revealed non-reciprocal directional dicroism, which is the light absorption difference for unpolarized counter-propagating  beams \cite{Tokura2007,Saito2008jpsj,Kezsmarki2011,Miyahara2011,Bordacs2012,Takahashi2013,Kezsmarki2014,Kezsmarki2015, Bordacs2015,Kuzmenko2015, Toyoda2015, Iguchi2017, Kocsis2018, Yu2018, Tokura2018, Kocsis2019,Viirok2019, Yokosuk2020, Kimura2020,Kimura2020PRL,Ogino2020,Kimura2021, Toyoda2021, Vit2021,Reschke2022}.
This unusual phenomenon is the finite-frequency manifestation of the magnetoelectric (ME) effect, which emerges at simultaneously electric- and magnetic-dipole allowed excitations, that we term as ME resonance\footnote{Usually, magnons couple to the magnetic component of electromagnetic radiation, i.e. they are magnetic-dipole active.
If the magnons are electric-dipole active, the term electromagnon is often used~\cite{Pimenov2006Nature}.
Magnetoelectric resonance is a \sw/ excited by both components of electromagnetic radiation~\cite{Kezsmarki2011,Takahashi2012}. 
For the rest of the paper we classify the \sw/s, based on their coupling to the electromagnetic radiation, using magnetic-dipole active, electric-dipole active and magnetoelectric \sw/. 
We use ``magnon'' for the \sw/ excitation described by the mean-field model without making a difference in its coupling to the electromagnetic radiation.}.
Since the relative orientation of the electric and magnetic fields is different for counter-propagating beams ME coupling generates an absorption difference  between the two beams and may even lead to one-way transparency \cite{Kezsmarki2014}. 
This non-reciprocal absorption may gain applications in photonics and spintronics~\cite{Kezsmarki2011,Kocsis2018}.
For example, materials with ME resonances can be used as optical diodes where the direction of transparency for the THz radiation can be switched by magnetic  fields~\cite{Miyahara2011,Kezsmarki2011,Bordacs2012,Kezsmarki2014,Bordacs2015,Kezsmarki2015,Viirok2019},  electric fields~\cite{Kuzmenko2018,Kimura2020PRL,Vit2021}, or both~\cite{Kocsis2018}.
From the fundamental science point of view, the spectroscopy of the ME excitations promotes the understanding of the static ME response which is linked to the non-reciprocal directional dichroism spectrum via the Kramers-Kronig relations~\cite{Szaller2014,Kocsis2019}.
Moreover, a THz absorption study, combined with magnetization, inelastic neutron scattering measurements~\cite{Li2006, Toft-Petersen2015, Yiu2017}, and theoretical modeling \cite{Szaller2017TFBO,Szaller2020,Room2020,Farkas2021} can resolve realistic spin Hamiltonians of ME compounds.

The relativistic spin-orbit coupling plays an essential role for ME spin excitations. 
It establishes a coupling between spins and electric dipoles and also introduces single-ion anisotropy for $S>1/2$.
The single-ion anisotropy expands the frequency scale of spin excitations as it separates the $\pm m_s$ doublets from each other in zero field, where $m_s$ is the spin quantum number.	
In addition to conventional spin waves, spin-quadrupolar excitations corresponding to $\Delta m_s =\pm 2$ may appear in systems with strong single-ion anisotropy and spin $S>1/2$~\cite{Penc2012,Romhanyi2012,Akaki2017,Legros2021,Bai2021}, broadening the frequency range for possible applications of ME materials.
In general, if there are $N$ spins in the magnetic unit cell we expect $2 N S$ spin excitations, which can be described by the multi-boson \sw/ theory~\cite{Penc2012,Romhanyi2012,Kocsis2018,SWBook2018} or by single-ion spin Hamiltonian with added molecular field to take into account spin-spin interactions~\cite{Chaix2014,Yiu2017,Strinic2020}.

\begin{figure}
		\centering
		\vspace{-8pt}
		\includegraphics[width=0.92\linewidth]{{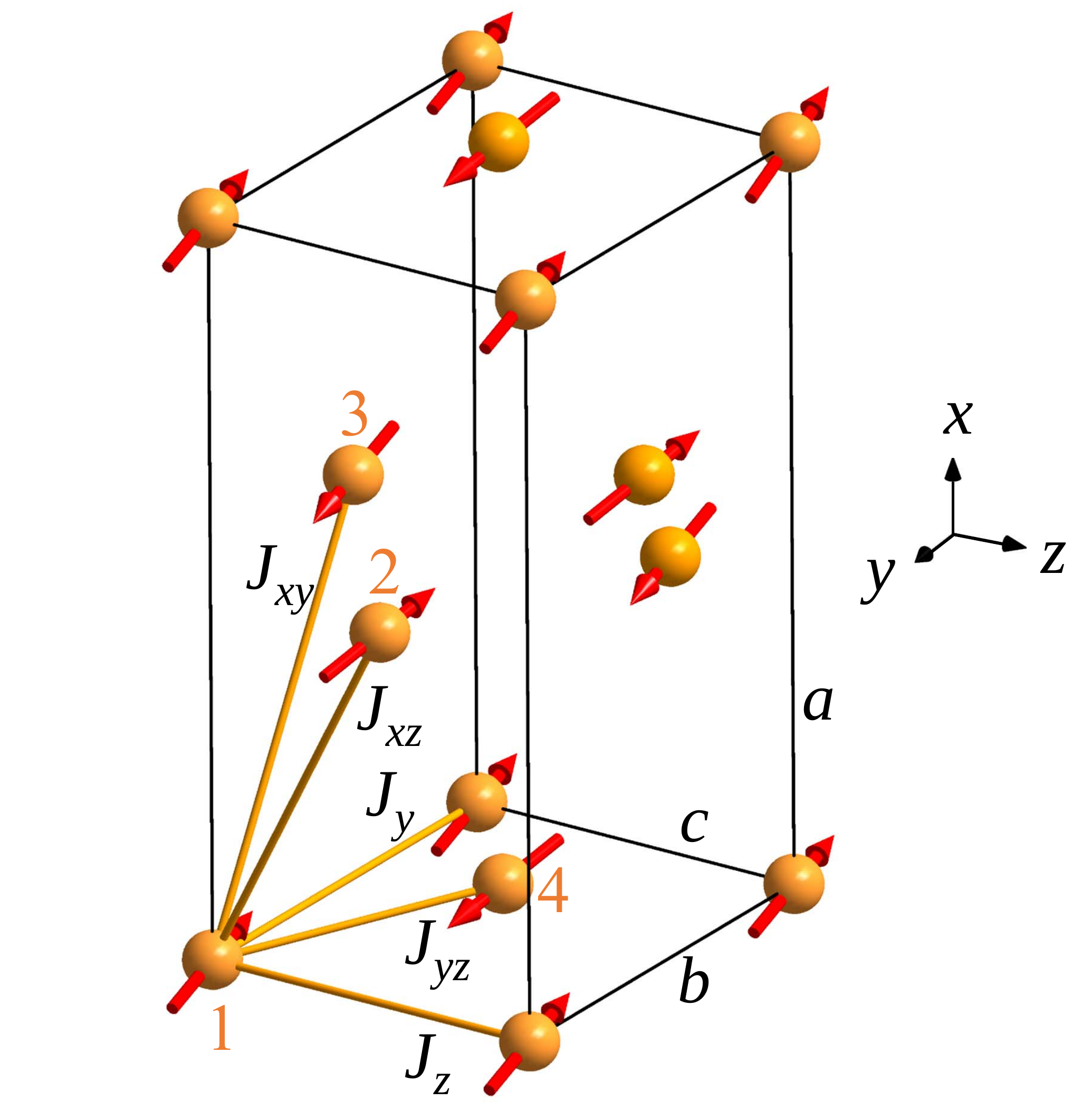}}
		\caption{\label{fig:magnetic_cell}
			The ground state spin configuration of \lfpo/  in  zero magnetic field. 
			There are four Fe$^{2+}$  spins, $S=2$,  in the magnetic unit cell drawn as a box. 
			$S_1$ is in the corner, $S_2$ is on the face and $S_3$ and $S_4$ are inside the unit cell.
			The spins are rotated towards the  $x$ axis and canted towards the $z$ axis away  from the $y$ axis \cite{Toft-Petersen2015}.
			The numbering of spins and the labeling of exchange interactions corresponds to the spin Hamiltonian described by Eq.\,(\ref{equation:model}).
		}
	\end{figure}

The \lmpo/ ($M=$ Mn, Co, Fe, Ni) orthophosphate compounds become ME as their magnetic order breaks the inversion symmetry~\cite{Santoro1967}.
This, together with their large single-ion anisotropy~\cite{Toft-Petersen2015,Yiu2017,Werner2021}, makes them appealing candidates to explore unconventional spin excitations.
Among these compounds, \lfpo/ has the highest N\'eel temperature, \tn/\,\,$=50$\,K below which an antiferromagnetic (AFM) order develops, as depicted in Fig.~\ref{fig:magnetic_cell}. 
The spins of the four magnetic ions of the unit cell are nearly parallel to the $y$ axis \cite{Creer1970}. 
Detailed neutron diffraction experiments showed that the spins are slightly rotated in the $xy$ plane and canted toward the $z$ axis \cite{Toft-Petersen2015}.
\lfpo/ has one of the largest spins in the orthophosphate family but  the number of \sw/ modes detected in the magnetically ordered phase has been substantially less than $2NS=16$, allowed for a  $S=2$  spin system.
In zero-field inelastic neutron scattering  (INS) studies two \sw/ branches~\cite{Li2006, Toft-Petersen2015, Yiu2017} and a dispersionless mode were observed below 10 meV~\cite{Yiu2017}. 
Whereas, a recent high-frequency electron spin resonance study detected two modes in the vicinity of the spin-flop field,  32\,T~\cite{Werner2021}.
Therefore, further experimental data is needed to understand better the spin dynamics and spin Hamiltonian of \lfpo/.

In this work, we studied the magnetic field dependence of the spin excitations using THz absorption spectroscopy in the low temperature AFM phase of \lfpo/.
The spectral range of our experiments extending up to 175\,\wn/ (22\,meV) covers two and five times larger energy window compared to former INS \cite{Li2006,Toft-Petersen2015,Yiu2017} and electron spin resonance studies \cite{Werner2020,Werner2021}, respectively.
The broader spectral range allowed us to observe 17 spin excitations and to determine their selection rules.
The absorption  spectra were measured with magnetic field applied along all three principal crystallographic axes up to 33\,T in the Faraday configuration (light propagates along the field,   $\mathbf{k}\!\parallel\!\hdc/$) and up to 17\,T in Voigt geometry (light propagates perpendicular to the field, $\mathbf{k}\!\perp\!\hdc/$).
Beside THz spectroscopy, we measured high-field magnetization up to 120\,T along the easy-axis from which we determined the spin-flop and the saturation fields.
Finally, we successfully employed a mean-field model to describe the field dependence of the magnetization and the resonance frequencies of the four strongest low-frequency spin-wave modes in the AFM state.
By fitting the magnetic field dependence of  four magnons we have refined the values of the exchange couplings, the single-ion anisotropies, and the \DM/ interaction.

	\section{Experimental}
	
	The \lfpo/ single crystals were grown by the floating zone method [\onlinecite{Baker2011}].
	The quality of the crystals was verified by powder diffraction and Laue XRD, which confirmed the orthorhombic structure with the same lattice constants as reported in Ref.~[\onlinecite{Garcia-Moreno2001}].
	
	The low field magnetization measurements were done using a 14\,T PPMS with VSM option (Quantum Design).
	High-field magnetization measurements were carried out up to 120\,T using ultra-high semidestructive pulses at the Laboratoire National des Champs Magn\'etiques Intenses in Toulouse~\cite{Portugall1999,Takeyama2012}.
	The maximum field of a semidestructive pulse was reached in $\sim$2.5\,$\mu$s.
	
	For THz spectroscopy studies the single crystal  was cut into three 1\,mm thick slabs each with a large face normal to one of the  three principal crystallographic directions.
	The slabs were wedged by two degrees to suppress the fringes in the spectra produced by the internal reflections in the crystal.
	
	The THz measurements up to 17\,T were performed  with a polarizing Martin-Puplett interferometer and a 0.3\,K silicon bolometer in Tallinn.
	High field spectra from 17\,T up to 33\,T were measured  using a Bruker IFS\,113v spectrometer and a 1.6\,K silicon bolometer in High Field Magnet Laboratory in Nijmegen.
	The experiments above 17\,T were done in Faraday configuration, while below 17\,T both Faraday and Voigt configuration experiments were performed.
	All spectra were measured with an apodized spectral resolution of 0.3 or 0.5\,\wn/.
	The polarizer angle with respect to the crystal axes was determined by evaluating the intensity change of the strongest modes in the THz absorption spectra as the function of rotation angle of the polarizer. 
	This information was also used to mount the polarizer in the high field experiments in Nijmegen where the in situ polarizer rotation was not possible. 
	Absorption was determined by using a reference spectrum of an open hole, sample spectrum in the paramagnetic phase or by statistically calculating the baseline from the magnetic field dependence of sample spectra.
	In the first method, the  absorption was calculated as
	\begin{equation}
	    \alpha = -d^{-1} \ln(I/I_r),
	\end{equation}
	where  $I_{r}$ is the intensity through the reference hole with the area equal to the sample hole area and $d$ is the sample thickness. 	
	In the second method, the absorption difference was calculated,
	\begin{eqnarray}
	\Delta\alpha(H,T) &=& \alpha(H,T) - \alpha(0\,\text{T},55\,\text{K})\nonumber\\
	&=&  -d^{-1} \ln\left[I(H,\textit{T})/I(0\,\text{T},55\,\text{K}) \right],
	\end{eqnarray}
	where   $I(0\,\text{T},55\,\text{K})$ is the intensity  through the sample measured at 0\,T and 55\,K in the paramagnetic phase.
In the third method, the statistically calculated baseline, $\alpha(0\,\text{T})$, 
was   found   as a minimum of differential absorption,
\begin{eqnarray}
   	\Delta\alpha_H(H_i)&=& \alpha(H_i)- \alpha(0\,\text{T}) \nonumber\\
   	&=& -d^{-1} \ln\left[I(H_i)/I(0\,\text{T} \right],
\end{eqnarray} 
at each frequency over several   magnetic field values $H_i$.
	By adding   $\alpha(0\,\text{T})$ to the differential absorption we get  the dependence of absorption spectra  on magnetic field.
	This method was used to obtain the spectra measured above 17\,T.
		
	\begin{center}
		\begin{figure}
			
			\includegraphics[width=8truecm]{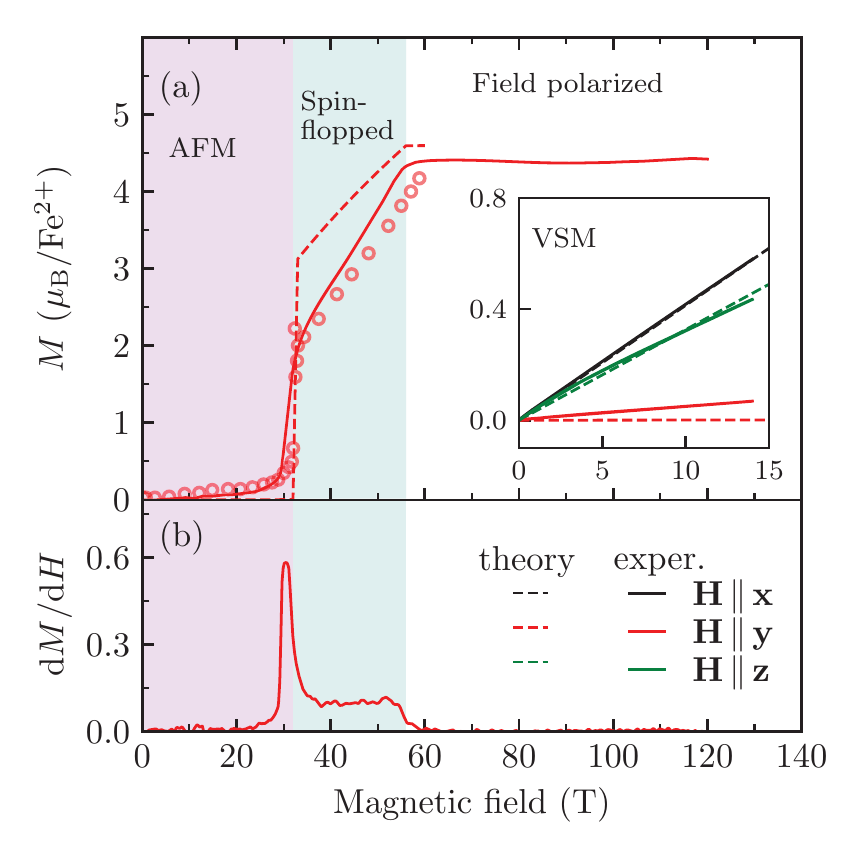}
			\caption{\label{fig:LFPO_magnetization}
		(a) Magnetic field dependence of the magnetization $M$ (solid, red) and (b) the directly measured $\mathrm{d}M/\mathrm{d}H$ (solid, red) at $T=5\,\mathrm{K}$ for increasing pulsed field in \hparal{y}.
			The inset of panel (a) shows VSM $M$-$H$ measurements in quasi-static fields at $T=2.4\,\mathrm{K}$, where the magnetic field directions are \hparal{x} (black), \hparal{y} (red) and \hparal{z} (green).
			The dashed lines in panel (a) show the results of the mean-field calculations with the parameters from Table\,\ref{tab:exchange parameters}.
			For comparison, we show the pulsed field magnetization data from Ref.~\cite{Werner2021} with open red circles.
			The AFM, spin-flopped and spin polarized state regions are shown for \hparal{y}.
				}
		\end{figure}
	\end{center}
	
	\begin{center}
	\begin{figure*}[htp] 				
		\includegraphics[width=17.5truecm]{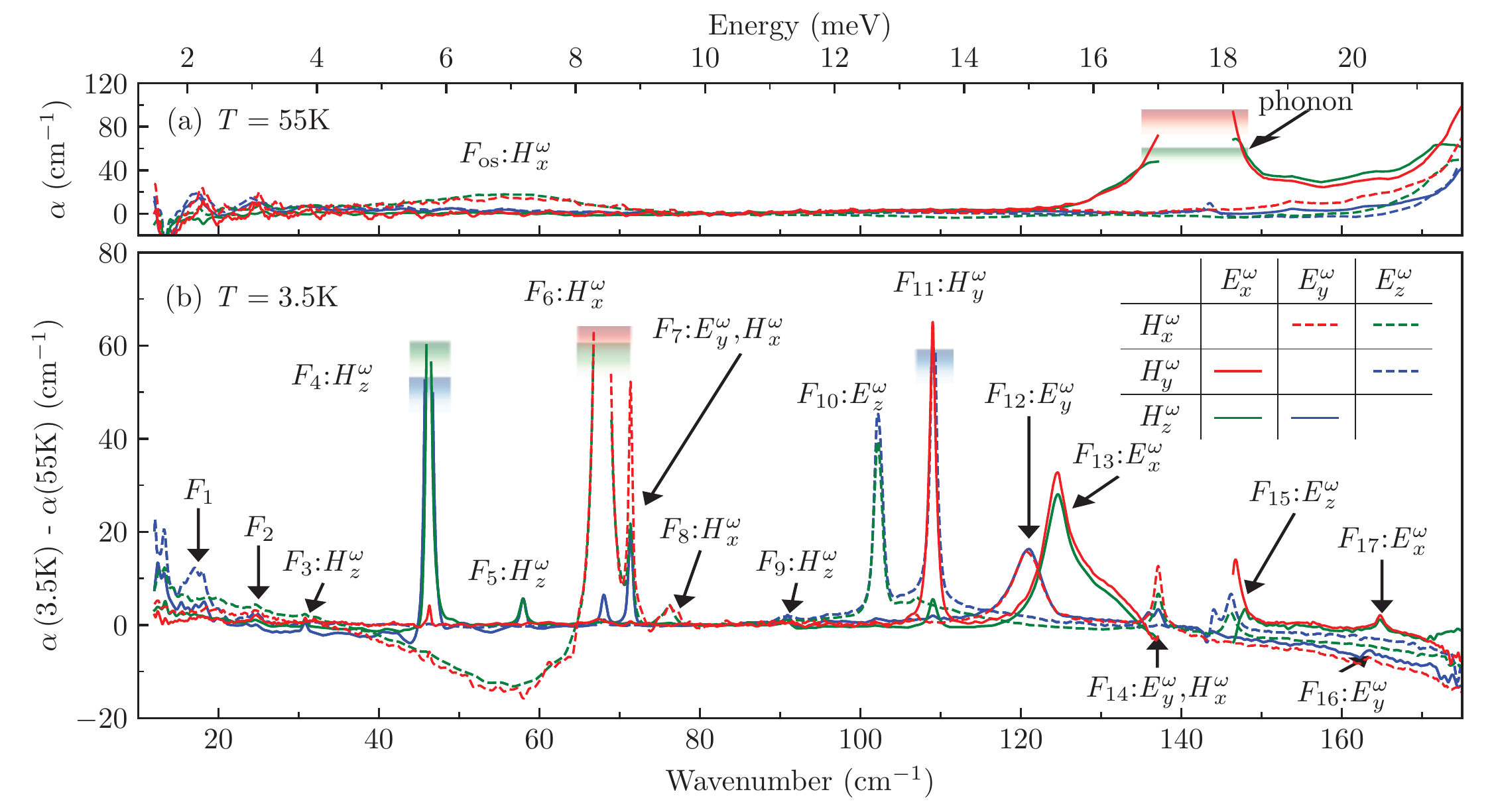}
		\caption{\label{fig:LFPO_selection_rules} (a) THz absorption spectra of \lfpo/ at $55$\,K in the paramagnetic phase, and (b) the difference between the zero-field absorption spectra recorded at  $3.5$\,K (magnetically ordered phase) and 55\,K, demonstrating spectral features associated with the onset of magnetic order.
		Line colors correspond to the propagation direction of the THz radiation: $k_x$ (blue), $k_y$ (green) and $k_z$ (red). 
		Two orthogonal polarizations $\{E^\omega_i, H^\omega_j\}$ for the given  propagation direction, $k_k\sim E^\omega_i \times H^\omega_j$, are indicated by the solid and dashed lines, according to the inset of panel (b).
		$F_n$ with $n=1,\ldots, 17$ labels the modes in the magnetically ordered phase and \fos/ is an on-site magnetic excitation in the paramagnetic phase.
		$H^\omega_j$ or $E^\omega_i$ indicate the magnetic- or electric-dipole activity of the mode, respectively.
		The blue, green and red rectangles mark the peaks with absorption above the upper detection limit, \mode{4}, \mode{6}, \mode{11}, and phonon.
		}
	\end{figure*}
	\end{center}
	\section{Mean-field model}\label{sec:Mean-field model}
	
The mean-field theory of localized magnetic moments is a widely applied tool to interpret the static and dynamic magnetic properties of systems with periodic magnetic structures~\cite{Turov1963}, e.g. ferro-~\cite{Kittel1948}, ferri-~\cite{Szaller2020} and antiferromagnetic-~\cite{Szaller2017TFBO} insulators. 
Particularly, the microscopic spin Hamiltonian of \lfpo/ has been discussed by several papers~\cite{Li2006,Liang2008,Toft-Petersen2015,Yiu2017,Werner2019,Werner2021}.

Here we aim to describe the static magnetism and the infrared-active optical magnetic resonances of \lfpo/. 
Thus, we use a simplified Hamiltonian where the exchange coupling terms $J_y$ and $J_z$ have been omitted as they connect spins at crystallographically equivalent sites, see Fig.~\ref{fig:magnetic_cell}. 
While the one-magnon THz spectrum is insensitive to the same energy shift of all states at the $\Gamma$ point of the Brillouin zone produced by $J_y$ and $J_z$, these couplings are relevant when describing the dispersion of the magnon modes~\cite{Toft-Petersen2015}. 
The Hamiltonian of our study is
\begin{eqnarray}
	\mathcal{H}& =&
 4 \left[ J_{xz}\left(\mathbf{S}_{1}\cdot\mathbf{S}_{2}+\mathbf{S}_{3}\cdot\mathbf{S}_{4}\right)\right. \nonumber\\ 
	&&+ J_{xy}\left(\mathbf{S}_{1}\cdot\mathbf{S}_{3}+\mathbf{S}_{2}\cdot\mathbf{S}_{4} \right) \nonumber \\
	&&+  J_{yz}\left(\mathbf{S}_{1}\cdot\mathbf{S}_{4}+\mathbf{S}_{2}\cdot\mathbf{S}_{3}\right)\nonumber \\
	&&+ \left.  D_{14}\left(S_1^y S_4^z - S_1^z S_4^y +  S_3^y S_2^z -  S_3^z S_2^y  \right)  \right]\nonumber\\
	&+& \sum_{i=1}^{4} \left[\Lambda_x\left(\vphantom{S_i^y}S_i^x\right)^2 + \Lambda_z\left(S_i^z\right)^2 + \Lambda_{xy}\left(S_i^x S_i^y \right)\right. \nonumber \\
	&& - \left. \mu_{\mathrm B}\mu_{0}\left( g_x\mathit{H}_x\mathit{S}^x_i+g_y\mathit{H}_y\mathit{S}^y_i+g_z\mathit{H}_z\mathit{S}^z_i\right)\right],
	\label{equation:model}
\end{eqnarray}
where the terms, exchange interactions, \DM/ term, single-ion anisotropy terms, and the Zeeman energy, have been considered in the earlier works of \lfpo/ summarized in Table\,\ref{tab:exchange parameters}.
The model is based on four Fe$^{2+}$ spins, here represented by classical vectors of $S\,=\,2$ length, that occupy crystallographically non-equivalent positions of the unit cell.
As shown in Fig.\ref{fig:magnetic_cell}, the  spins are coupled by three different exchange couplings with parameters $J_{xz}, J_{xy}$ and  $J_{yz}$.
There are two single-site hard-axis anisotropies, $\Lambda_x$ and $\Lambda_z$, that effectively produce the easy axis along $y$.
The spins are slightly rotated away from the $y$ axis towards the $x$ axis as observed by neutron scattering~\cite{Toft-Petersen2015}. 
Extending previous studies to reproduce this small deviation of the magnetic structure from the collinear antiferromagnetic order we introduced an additional single-site anisotropy term $\Lambda_{xy}S^x S^y$.
The \DM/ interaction is $\vect{D}=\{D_{14},0,0\}$.
Since the spins are predominantly along the $y$ direction,  $D_{14}$ cants  spins towards the  $z$ direction.
The last term in Eq.\,(\ref{equation:model}) is the interaction of the electron spin with the applied magnetic field taking into account the $g$-factor anisotropy.

We model the spin dynamics using  the Landau-Lifshitz-Gilbert equation, Ref.\,\cite{Gilbert2004}, as used in Ref.~\cite{Peedu2019}, by assuming  that the spins are oscillating about their equilibrium orientations without changing their lengths.
The equilibrium orientation of the spins is  found by minimizing the energy described in the Hamiltonian of  Eq.\,(\ref{equation:model}), with respect to the spin orientations.
The magnetic-dipole absorption of light by \sw/s is calculated assuming that the magnetic field $\vect{H}^\omega$ of  radiation  couples to the total magnetic moment of the spins~\cite{Peedu2019}.
Dielectric permittivity in the absorption coefficient formula, Eq.\,(10) in Ref.\,\cite{Peedu2019}, was assumed to be real  and frequency-independent with components $\epsilon_{x}=8.1$,   $\epsilon_{y}=7.3$ and $\epsilon_{z}=7.6$~\cite{Stanislavchuk2017}.

	\def\arraystretch{1.5}	
	\begin{table}		
		\begin{ruledtabular}	
			
			\caption{The parameters of the mean-field model used  to describe the static magnetic properties and  \sw/s in \lfpo/:
				exchange couplings $J_{i}$ and $J_{ij}$, single-ion anisotropies $\Lambda_{i}$ and $\Lambda_{ij}$, \DM/  coupling $D_{14}$, and anisotropic  $g$-factor $g_i$. 
				All parameters are in units of meV except the dimensionless $g_i$.
			} \label{tab:exchange parameters}	
			\vspace{5pt} 
			\begin{tabular}{ *{9}{c} }
			$J_{xz}$ & $J_{xy}$ & $J_{yz}$ & $\Lambda_{x}$ & $\Lambda_{z}$ & $\Lambda_{xy}$ & $D_{14}$ & $g$ & Ref. \\ \hline
				$-0.006$ & 0.086 & 0.51 & 0.52 & 1.52 & $-0.009$ & 0.027 & $g_x=2.04$ & \footnote{This work} \\[-0.5em]
				  &  &  &  &  &  &  & $g_y=2.3$ &\\ [-0.5em]
				   &  &  &  &  &  &  & $g_z=2.1$ &\\
			 \hline 
				 0.05 & 0.14 & 0.77 & 0.62 & 1.56 & - & 0.038\footnote{The \DM/ parameter $D_{14}=J_\mathrm{DM}/4 $ where $J_\mathrm{DM}$ is from Ref.\,\cite{Werner2021}.} & $g_x=2.24$ & [\onlinecite{Werner2021}]  \\[-0.5em]
				  &  &  &  &  &  &  & $g_y=2.31$ &\\ [-0.5em]
				   &  &  &  &  &  &  & $g_z=1.99$ &\\ \hline 
				 0.01 & 0.09 & 0.46 & 0.86 & 2.23 & - & - &- &[\onlinecite{Yiu2017}]  \\ \hline
				0.05 & 0.14 & 0.77 & 0.62 & 1.56 & - & - & -&[\onlinecite{Toft-Petersen2015}]  \\
				
			\end{tabular}
		\end{ruledtabular}
	\end{table}

	\section{Results}
	
\subsection{Magnetization}
	We characterized \lfpo/ samples by measuring the magnetization at 2.4\,K along the principal axes up to 14\,T. Along \hparal{y}, the measurement is extended up to 120\,T at 5\,K using pulsed fields, see Fig.~\ref{fig:LFPO_magnetization}.
	The $y$-axis magnetization determined from the pulsed-field measurements was normalized to the value of static field measurements in the range from 4 to 14\,T, neglecting a small hysteresis of magnetization between 0 and 4\,T.
	In the AFM state the spins are predominantly aligned along the easy axis, the $y$ axis in \lfpo/.
	The magnetization grows approximately linearly in increasing field for \hparal{x} and \hparal{z}.
	When \hparal{y} is applied, the spins maintain easy-axis alignment up to the spin-flop field marked by a jump in the magnetization at 32$\pm$3\,T.
	As the field further increases the magnetization grows linearly and reaches saturation at 56$\pm$3\,T. 
	In the field-polarized state, the saturation magnetization is estimated to 4.4\,$\mu_B$ per iron.
	The spin-flop field deduced from our measurements is in agreement with former high-field magnetization measurements \cite{Werner2021}.

\subsection{THz absorption spectra in zero field}	
	The zero-field THz absorption spectra of \lfpo/ are presented in Fig.~\ref{fig:LFPO_selection_rules} and the mode parameters are collected in Table~\ref{tab:LFPO_modes}, while Fig.~\ref{fig:LFPO_selection_rules}(b) features absorption spectra in the AFM phase, relative to the paramagnetic phase.
	
	The spectra in the paramagnetic phase show  a broad but weak magnetic-dipole active peak \fos/ at around $\sim$55\,\wn/, Fig.~\ref{fig:LFPO_selection_rules}\,(a).
	The magnetic on-site excitation \fos/ is \hwu{x} active as it is seen in two polarization configurations, \{\ewu{y},\hwu{x}\} and \{\ewu{z},\hwu{x}\}.
	The frequency  and the selection rules of \fos/  are reproduced  by exact diagonalization of a four-spin cluster, see Fig.\,S6 in the Supplementary Material.
	Other features in the paramagnetic phase spectra are 
\ewu{x}-active phonon at 140\wn/, with a strong absorption exceeding the detection limit, and
absorption rising towards higher frequencies  due to the phonons with resonance frequencies  above 175\wn/. 
	
	To better resolve spectral features emerging in the magnetically ordered phase we plot the difference spectra, $\alpha(3.5\,\mathrm{K})-\alpha(55\,\mathrm{K})$,  Fig.~\ref{fig:LFPO_selection_rules}\,(b).
	We observe a diminished absorption at the tails of phonons at low $T$ appearing as negative features in the difference spectra between 140 and 175\wn/.
	The change of the 140\wn/ phonon mode is obscured by the strong absorption and therefore the \ewu{x}-spectra, green and red solid lines, are discontinued where the 140\wn/ phonon peaks.
	The  broad   peak \fos/ from the high-$T$ paramagnetic phase appears as a negative feature in the difference spectra in \hwu{x} polarization.
	
	All sharp modes, labeled \mode{1},$\ldots$, \mode{17}, are absent above \tn/ and we assign them  to spin excitations.
	The seven excitations, \mode{3}, \mode{4}, \mode{5}, \mode{6}, \mode{8}, \mode{9} and \mode{11}, are identified as magnetic-dipole active modes.
	Six modes, \mode{10}, \mode{12}, \mode{13}, \mode{15}, \mode{16} and \mode{17}, are identified as electric-dipole active resonances.
	The  mode \mode{13} has a shoulder, thus, it was fitted with two Gaussian lines with maxima at 124.4\wn/ and 127.6\wn/.
	Two modes, \mode{7} at  71.4\wn/ and \mode{14} at  137.1\wn/, are both electric- and magnetic-dipole allowed, therefore, we identified them as ME resonances.
	\mode{7} is the strongest in \{\ewu{y},\hwu{x}\} polarization, red dashed line in Fig.\,\ref{fig:LFPO_selection_rules}\,(b), and its intensity is halved if only one of the components, \ewu{y} or \hwu{x},  is present.
	Thus, \mode{7} is an example of a ME resonance which couples equally to the magnetic and electric components of radiation.
    We detected \mode{14} in the same three polarization configuration, thus, we also assigned it to a ME resonance with the same selection rule as mode \mode{7}, \{\ewu{y},\hwu{x}\}.
	
	The three strongest magnetic-dipole active modes \mode{4}, \mode{6}, and \mode{11} show only weak absorption in polarizations orthogonal to their main magnetic dipole component.
	The weak absorption in other polarizations could be explained by the imperfections of the polarizer.
	However, we can not completely rule out that some of these modes are ME resonances with a weak electric-dipole component which can be tested by further measurements of the non-reciprocal directional dichroism on magneto-electrically poled samples~\cite{Kocsis2018,Kocsis2019}.
	We can not identify the selection rules for modes \mode{1} and \mode{2} as they are too weak.

\subsection{Magnetic field dependence of \sw/s}	
	
\begin{center}
		\begin{figure*}	
			\includegraphics[width=16truecm]{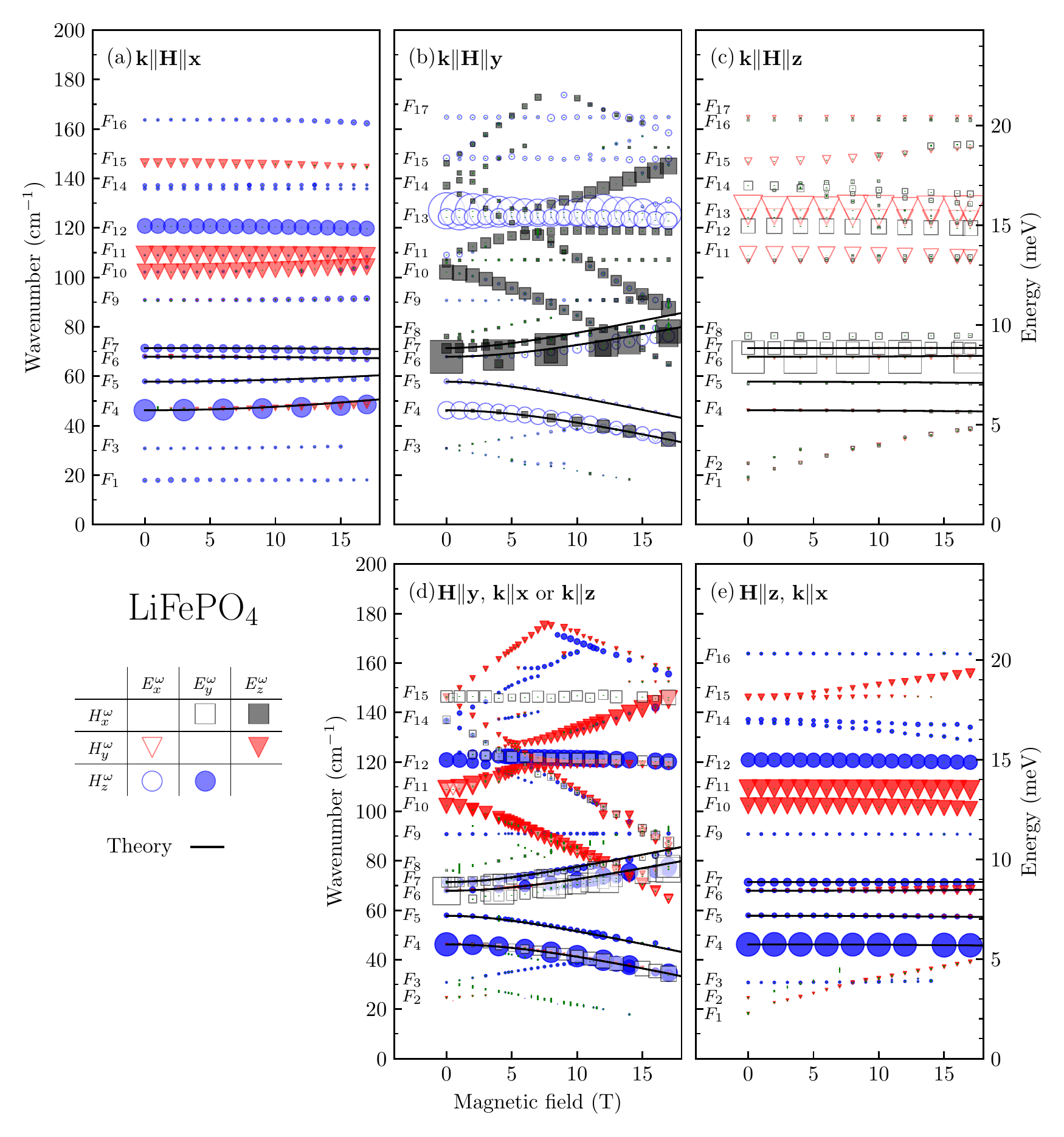}
			\caption{\label{fig:lfpo_5x_panels_scatter_plot}
			 	Magnetic field dependence of the \sw/ resonance frequencies and absorption line areas at $T=3.5\,\mathrm{K}$ in 	\lfpo/. 
				Panels (a), (b), and (c) correspond to measurements in the Faraday  ($\kvec/\parallel \hdc/$), while panels (d) and (e) correspond to experiments in the Voigt ($\kvec/\perp \hdc/$) configuration.
				The direction of the magnetic field is (a) -- \hparal{x}, (b), (d) -- \hparal{y}, and (c), (e) -- \hparal{z}.
				The symbols correspond to six combinations of linear light polarization $\{E^\omega_i, H^\omega_j\}$ as indicated at bottom left of the figure.
				The symbol height is proportional to the square root of experimental absorption line area with the same scaling as wavenumber axis.
				To simplify the figure the larger symbols are not shown for every measured field. 
				The error bars (vertical green lines) from fitting the line positions  in most cases are too small to be seen in the figure.
				The black lines are the results of the mean-field model calculations, modes \mode{4}, \mode{5}, \mode{6}, and \mode{7}.
				Comparison of   experimental and calculated intensities is in the Supplementary Material.
			}
		\end{figure*}	
	\end{center}

	The magnetic field dependence of mode frequencies and intensities between 0 and 17\,T is shown in Fig.\,\ref{fig:lfpo_5x_panels_scatter_plot} for Faraday, panels (a)-(c), and Voigt configuration,   (d) and (e).
	The modes mostly stay at constant frequency when the magnetic field is applied along the hard  axes, \hparal{x}, Fig.\,\ref{fig:lfpo_5x_panels_scatter_plot}(a) and \hparal{z}, Fig.\,\ref{fig:lfpo_5x_panels_scatter_plot}(c, e).
	However, most of the resonances shift with the magnetic field for \hparal{y}.
	We assigned a slope, $\slope/ = \Delta E/\Delta B$, calculated between 15 and 17\,T in units \wn/T$^{-1}$, to each of the modes  and collected them in Table\,\ref{tab:LFPO_modes}. 
	If the mode was not visible in this  range, a lower magnetic field range was used.
	One mode, \mode{17}, has zero slope and \mode{9}, \mode{13}, and \mode{16} have a moderate value, $|\slope/|<0.3$.
	Modes \mode{14} and \mode{15} have the largest $|\slope/|$ for \hparal{y} but also a substantial $|\slope/|$ for \hparal{z}.
	
	Assuming $g\approx 2$ we estimated from the slopes the change of the spin projection quantum number, $\Delta m_s$, upon the excitations.
	The results are listed in Table\,\ref{tab:LFPO_modes}. 
	The \sw/s below 80\wn/ (zero-field frequency) have $|\Delta m_s|=1$ while above 100\wn/ $|\Delta m_s|$ is  2, 3 or 4.
	 $|\Delta m_s|$ was not assigned to \mode{1} and \mode{2} where $\slope/\approx 1.5$\wn/T$^{-1}$ below 8\,T, which is between $\Delta m_s=1$ and 2.
	 We note that  \slope/ of \mode{1} changes with field.
	 It  is 0.9\wn/T$^{-1}$ above 8\,T. 
	 This change of slope could be due to the anti-crossing with \mode{4} but we do not have evidence for that because the mode was too weak to be detected in the high-field magnet set-up above 17\,T.
	
\begin{center}
	\begin{figure*}
		\includegraphics[width=\textwidth]{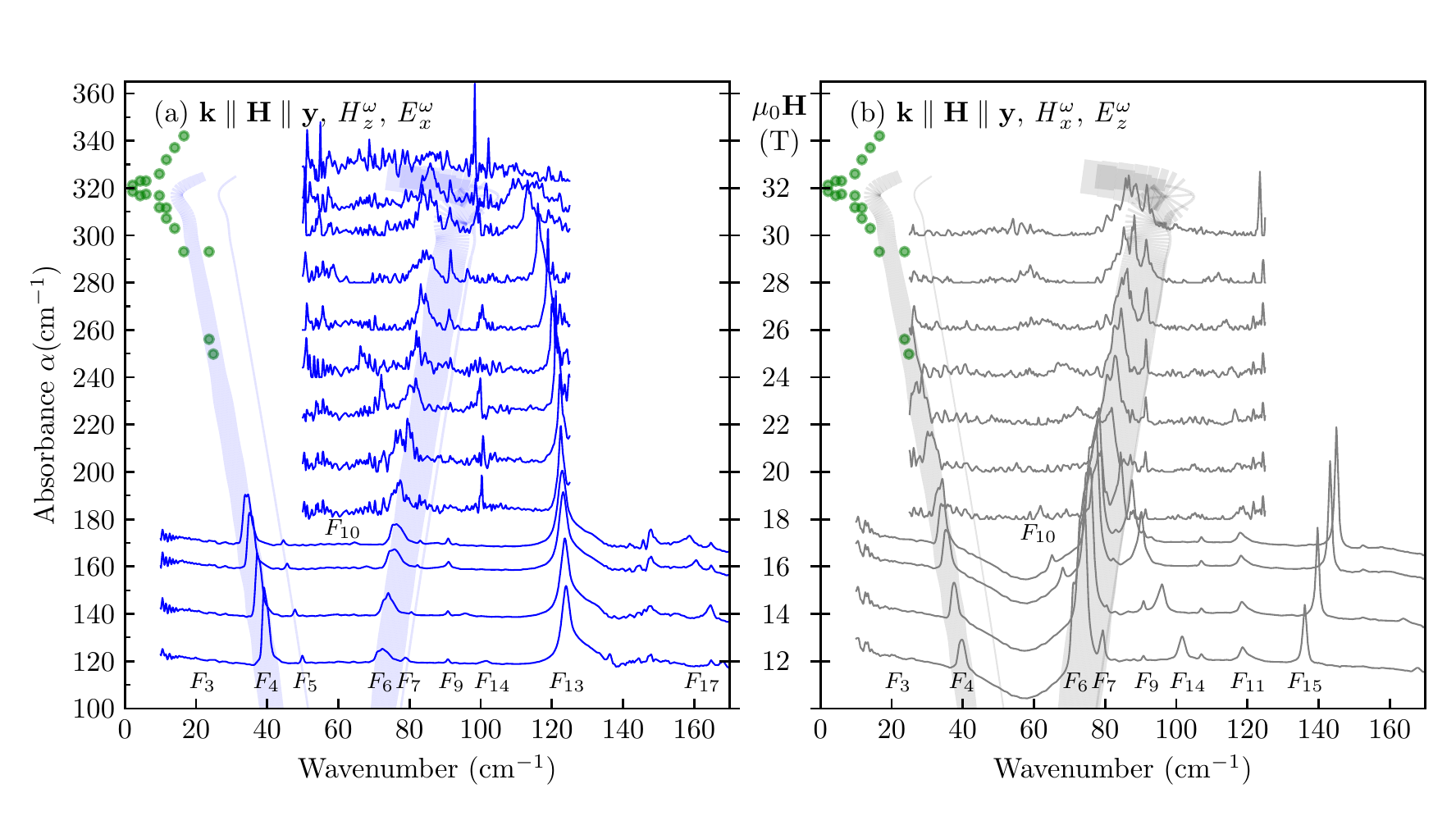}
		\caption{\label{fig:LFPO_H[y]_Faraday_two_panel}
			THz absorption spectra of \sw/ excitations in magnetic field $\vect{H} \parallel \vect{k}\parallel\vect{y}$ at $T = 3.5$\,K in two orthogonal polarizations, (a) $\{ \hwu{z},\ewu{x}\}$ and  (b) $\{ \hwu{x},\ewu{z}\}$.
			The broad stripes are the results of the mean-field  calculation, \mode{4}, \mode{5}, \mode{6} and \mode{7}, in order of increasing frequency;  the width of the line is proportional to the square root of the line area with the same scale as wavenumber axis and calculated in the magnetic dipole approximation.
			The zero field 55\,K spectrum was used as a reference below 17\,T and the low temperature zero-field spectrum above 17\,T.
			Filled circles are the \sw/ excitation energies reproduced from Ref.~\onlinecite{Werner2021}.
		}
	\end{figure*}
\end{center}

	The absorption spectra  in high magnetic field \hparal{y} up to  31.6\,T are presented in Fig.\,\ref{fig:LFPO_H[y]_Faraday_two_panel}.
	The \sw/  excitations, \mode{6} and \mode{7}, start softening before reaching the spin-flop transition at 32\,T, in accordance with the mean-field calculation.
	 Also, \mode{13} at about 125\wn/ shows softening close to 30\,T.
	 Spectra in other two field directions, \hparal{x} and \hparal{z}, above 17\,T  are shown in Supplementary Material, Figs.\,S1 and S3.

	\begin{table*}
		\centering
		\caption{\label{tab:LFPO_modes} The excitation configurations and field dependence of \lfpo/ modes in the AFM phase. 
			The selection rules were found by measuring polarization dependence of spin excitations in three principal directions without magnetic field.
			The absorption line energy and area in zero field were obtained from the fit to Gaussian  lineshape, except \mode{13} where the sum of two Gaussians was used. 
			The slopes of the modes  were estimated  from the  linear  field dependence  between 15 and 17\,T; if mode was not visible in this field range, the lower field range was used.
			From the slopes   the $|\Delta m_s|$ values are proposed assuming $g\approx 2$.
			Modes \mode{4} to \mode{7} were observed by INS spectroscopy~\cite{Toft-Petersen2015,Yiu2017} and are fitted to the mean-field model in this work.
			}

		\begin{tabular}{c d c c c c c c} 
			\hline\hline
			Mode &\multicolumn{1}{c}{Energy at 0\,T} & Area at 0\,T  &  Selection & \multicolumn{1}{c}{Magnetic field}&Slope \slope/ & $|\Delta m_s|$ \\
			 &\multicolumn{1}{c}{(\wnnospace/)} & (cm$^{-2}$) & rules at 0\,T & direction & (\wnnospace/T$^{-1}$) & \\
			\hline
			\mode{1} & 18.3 & 4 &   &  $z$ & $+1.4$ & \\ 
			\mode{2} & 24.7 & 2 &   &  $z$ & $+1.5$ & \\ 
			\mode{3} & 30.8 & 2 & \hwu{z} &   $y$ & $-0.9$, $+0.9$ & 1 \\ 
			\mode{4} & 46.2\,\,(5.7\,\mathrm{meV}) & $>$100 & \hwu{z} &  $y$ & $-1.1$ & $1$ \\
			\mode{5} & 58.0\,\,(7.2\,\mathrm{meV}) & 6 & \hwu{z} & $y$ & $-1.1$ & $1$ \\
			\mode{6} & 67.9\,\,(8.4\,\mathrm{meV})& $>$200 & \hwu{x} &  $y$ & $+0.9$ & $1$ \\
			\mode{7} & 71.4\,\,(8.9\,\mathrm{meV}) & 37 & \hwu{x},\,\ewu{y}&  $y$ & $+1.0$ & 1 \\
			\mode{8} & 76.2 & 9 & \hwu{x} &  $y$ & $-0.8$, $+1.0$ & 1 \\
			\mode{9} & 90.8 & 2 & \hwu{z} &  $x$ & $+0.1$ & \\
			\mode{10} & 102.2 & 57 & \ewu{z} & $y$ & $-3.3$ & 3\\
			\mode{11} & 109.0 & 74 & \hwu{y} & $y$ & $+1.8$ & $2$ \\
			\mode{12} & 120.8 & 50 & \ewu{y} & $y$ & $-1.9$ & $2$ \\
			\mode{13} &\multicolumn{1}{c}{ 124.4, 127.6} & 185 & \ewu{x} & $y$ & $-0.3$ & \\
			\mode{14} & 137.1 & 17 & \hwu{x},\,\ewu{y} & $y$ &  $-3.0$, $+2.8$ & $3$ \\
		 			  & & & & $z$ &  $-0.6$ & \\
			\mode{15} & 146.3 & 30 & \ewu{z} & $y$ & $-3.7$, $+3.8$  & $4$ \\
					  & & & & $z$ & $+0.7$  &  \\ 
			\mode{16} & 163.7 & 2 & \ewu{y} & $x$ & $-0.3$ & \\
			\mode{17} & 164.8 & 4 & \ewu{x} & $y$ & $0.0$ & \\
			\hline\hline
		\end{tabular}
	\end{table*}
	

\subsection{Mean-field model results}

The mean-field model parameters of Table\,\ref{tab:exchange parameters} were obtained by fitting the magnetic field dependence of frequencies of the \sw/ modes \mode{4}, \mode{5}, \mode{6}, and \mode{7}.
The magnetic field dependence of these modes below 17\,T is reproduced  remarkably well by the model for all three magnetic field directions, Fig.\,\ref{fig:LFPO_H[y]_Faraday_two_panel} and Fig.\,\ref{fig:lfpo_5x_panels_scatter_plot}.
The isotropic $g$-factor was not sufficient to quantitatively describe the magnetic field dependence of mode frequencies.
The anisotropic $g$-factor  values improved not only the magnetic field dependence of \sw/ frequencies, but also  reproduce the value of the spin-flop field and the saturation field, Fig.\,\ref{fig:LFPO_magnetization}.
In addition, the calculated magnetization as a function of $H_x$ and $H_z$ follows the measured  $M(H)$ below 15\,T, inset to Fig.\,\ref{fig:LFPO_magnetization}.

The equilibrium spin configuration deviates in two ways from the perfect collinear arrangement of spins along the $y$ axis.
With the parameters from Table\,\ref{tab:exchange parameters} we get the canting away from the $y$ axis towards the $z$ axis by 0.86 degrees, driven by the \DM/ interaction $D_{14}$.
The rotation of spins, driven by $\Lambda_{xy}$, away from the $y$ axis towards the $x$ axis is 0.95 degrees.
Using  the  spin length $S=2$, the out-of-easy-axis magnetic moments per spin are $|m_z|=0.063\mu_{\mathrm B}$  and $|m_x|=0.067\mu_{\mathrm B}$ reproducing  the experimentally determined deviations, $m_z$ = 0.063(5)$\mu_B$ and $m_x$ = 0.067(5)$\mu_B$~\cite{Toft-Petersen2015}.

The saturation value of the magnetization for \hparal{y} calculated from the mean-field model is 4.5\% higher than the experimentally observed, Fig.\,\ref{fig:LFPO_magnetization}\,(a).
Reason for the failure to reproduce the saturation magnetization and the spin-flop field with the
same set of magnetic-field independent parameters could be magnetostriction~\cite{Werner2019}.
Magnetostriction, as was proposed in Ref.\,\cite{Werner2021}, could also be the reason why the mean-field model does not reproduce the frequency of \mode{4} close to the spin-flop field, 32\,T in Fig.\,\ref{fig:LFPO_H[y]_Faraday_two_panel}.	

\section{Discussion}

\subsection{\Sw/ excitations from the mean-field model \label{Sec:Linear magnon excitations}}

We found that the the mean-field model quantitatively describes the magnetic field dependence of the frequencies of \sw/s \mode{4}, \mode{5}, \mode{6}, and \mode{7}, Fig.\,\ref{fig:lfpo_5x_panels_scatter_plot}. 
The modes \mode{4}, \mode{5}, \mode{6}, and \mode{7} have a linear field dependence with the slope close to $\pm 1$\wn/T$^{-1}$ when the field is along the easy axis $y$. 
This slope corresponds to a \sw/ excitation with $\Delta m_s=\pm1$, assuming $g\approx 2$. 
Other studies also found a g-factor close to 2~\cite{Werner2021}.
Other candidates for the $\Delta m_s=\pm1$ \sw/ excitations are \mode{3} and \mode{8}.
However, both of these modes have two branches degenerate in zero field. 
The magnetization measurements, inset of Fig.~\ref{fig:LFPO_magnetization}(a), indicate bi-axial magnetic anisotropy in \lfpo/ which lifts the degeneracy of magnetic resonances in zero field.
Therefore, \mode{3} and \mode{8} cannot be consistently included into the mean-field description.

The \sw/s of the mean-field model have oscillating spin components, $\delta \vect{S}_i=\vect{S}_i-\bar{\vect{S}}_i$, perpendicular to the equilibrium direction of the $i$-th spin, $\bar{\vect{S}}_i$.
The \sw/ couples to the magnetic field of radiation if the oscillating spin component of the whole magnetic unit cell is finite, 
$\vect{H}^\omega\cdot\left(\sum_{i=1}^{4}\delta \vect{S}_i \right)$. 
The equilibrium direction of the spins is aligned to the easy axis $y$ within few degrees in \lfpo/. 
The selection rules, Table\,\ref{tab:LFPO_modes}, show  that \mode{4} and \mode{5} are excited by the \hwu{z} component of radiation  and modes  \mode{6} and \mode{7} by the \hwu{x} component, which both are perpendicular to $\bar{\vect{S}}_i$.
The magnetic field dependence of intensities of the strongest modes \mode{4} and \mode{6}  is well described by the mean-field model.
Firstly, \mode{4} is \hwu{z}- and \mode{6} is \hwu{x} -active in zero field, Table\,\ref{tab:LFPO_modes}.
Secondly, as  $H_y$  increases, \mode{4} becomes \hwu{x}-active and \mode{6} becomes \hwu{z}-active, Fig.\,S4 in the Supplementary Material.
Thus, for modes \mode{4} and \mode{6} the agreement between theory and experiment is good.

The experimental and theoretical selection rules of magnetic dipole transition  for \mode{5} agree, it is \hwu{z}-active.
For the \sw/ \mode{7} the theory predicts \hwu{y}-activity, although it is \hwu{x}-active in the experiment, see Fig.\,S4 in the Supplementary Material.
Overall, theory underestimates \mode{5} and \mode{7} magnetic dipole transition intensity by two orders of magnitude.
It is not surprising as the modes \mode{5} and \mode{7} are relatively weak as compared to \mode{4} and \mode{6} and therefore they are sensitive to the composition of the \sw/ state. 
If the mean-field model, as an approximation, does not give the true \sw/  state, the weak intensities could be seriously affected.
As observed experimentally, Fig.\,\ref{fig:LFPO_selection_rules}, \mode{7} is  electric-dipole active  in addition. 
The coupling of spins to the electric field was not included in the mean-field model.

Similar to the quantum-mechanical formulas of the magnon dispersion relation of former studies~\cite{Li2006,Tian2008,Toft-Petersen2015}, the classical expressions for the zero-field resonance frequencies of the magnon modes can be derived. 
For the two strongest \sw/s \mode{4} and \mode{6}
\begin{equation}
\nu_{4/6}\approx 2S\sqrt{\Lambda_{x/z} \left( 4(J_{yz}+J_{xy}) + \Lambda_{z/x} \right)},
\label{eq:nu_46}
\end{equation}
while the zero-field frequencies of the weaker \mode{5} and \mode{7}  are
\begin{equation}
\nu_{5/7}\approx 2S\sqrt{\left( \frac{\Lambda_{x/z}}{1-\sqrt{J_{xy}/\Lambda_{x/z}}} -J_{xz}\right) \left( 4J_{yz} + \Lambda_{z/x} \right)},
\label{eq:nu_57}
\end{equation}
where we neglected the weak \DM/ interaction and  the  single-ion anisotropy $\Lambda_{xy}$ terms. 
While these two terms are necessary to give finite magnetic dipole activity to the weak \mode{5} and \mode{7} resonances by breaking the equivalence of $S_{1}$ and $S_{2}$ ($S_{3}$ and $S_{4}$, respectively) sublattices, they do not change the resonance frequencies significantly. 

As follows from Eq.~(\ref{eq:nu_46}) and Eq.~(\ref{eq:nu_57}), if $J_{xy}=J_{xz}=0$, \mode{4} and \mode{5} are degenerate in zero field, $\nu_4=\nu_5$,   and also $\nu_6=\nu_7$. 
In this case the nearest-neighbor $(100)$ planes  of the $\{ S_1,S_4\}$ and $\{ S_2,S_3\}$ sub-lattices, separated by $a/2$, are decoupled from each other, thus, their in-phase and out-of-phase excitations with respect to each other are degenerate. 
Consequently, \mode{4} and \mode{6} can be considered as the in-phase while \mode{5} and \mode{7} as the out-of-phase  resonances of the nearest-neighbor $(100)$ planes. 
Without \DM/ interaction and $\Lambda_{xz}$ anisotropy the total oscillating magnetic dipole moment of the unit cell produced by \mode{5} and \mode{7} is zero. 
This explains the weak intensity of \mode{5} and \mode{7} compared to \mode{4} and \mode{6} in the THz absorption spectrum. 
Furthermore, the correspondence between the INS magnon dispersion interpreted in the two-spin unit cell scheme~\cite{Li2006,Toft-Petersen2015,Yiu2017} and our $\Gamma$-point optical experiments can also be formulated based on the mean-field results. Namely, \mode{4} and \mode{6} correspond to the \sw/s observed in the zone center, $\vect{Q}=(0,2,0)$~\cite{Toft-Petersen2015} or $\vect{Q}=(0,0,2)$~\cite{Yiu2017} while \mode{5} and \mode{7} are zone-boundary excitations of the two-spin unit cell, seen at $\vect{Q}=(0,0,1)$~\cite{Yiu2017},  $\vect{Q}=(1,1,0)$~\cite{Li2006,Toft-Petersen2015} and $\vect{Q}=(0,1,1)$~\cite{Toft-Petersen2015} in the INS experiments~\cite{Li2006,Toft-Petersen2015,Yiu2017}.

\subsection{Spin excitations beyond the mean-field model \label{Sec:Exc_beyond_linear_spin_wave_approx}}
 		
Out of 17 lines appearing below \tn/ in the THz absorption spectrum only four can be described by the classical four-spin  mean-field model.
The rest can be (\emph{i}) spin-stretching excitations captured only by multi-boson spin-wave theory or alternatively by crystal-field schemes including exchange fields, (\emph{ii}) two-magnon excitations (two spin waves with nearly opposite \textbf{k} vectors), or can even be (\emph{iii}) excitations from impurity spins.
Assuming that the spins are aligned along the $y$-axis the magnetic symmetry reduces to $Pnma'$~\cite{Li2006}.
Since all spatial symmetries of the paramagnetic state remain in the AFM phase, at least in combination with time-reversal operation, we do not expect new optical phonon modes to emerge below \tn/. 

We assign absorption lines \mode{1}, \mode{2} and \mode{3} to impurities because  these very weak modes are located below the lowest magnon mode \mode{4}.
In addition, the frequencies of \mode{1} and \mode{2} increase linearly in magnetic field \hparal{z}, not coinciding with easy-axis direction $y$.
Previous works have found that Fe$^{2+}$ at Li$^{+}$ site has zero field splitting 7.3\wn/ (220\,GHz)~\cite{Werner2020}.
The lowest impurity absorption line in our spectrum is \mode{1} at 18\wn/ in zero field.
This suggests that we are observing different impurities than reported in Ref.~\cite{Werner2020}.

The mean-field model does not describe spin excitations \mode{8}\mbox{--}\mode{17}.
Several of them are electric-dipole active and have a steep magnetic field dependence of frequency, suggesting $|\Delta m_s|>1$ change of a spin projection quantum number.
This  is unusual for a \sw/ excitation but can be explained by a large single-ion anisotropy ($\Lambda$) which is comparable or stronger than the exchange coupling ($J$)~\cite{Penc2012}, see Table\,\ref{tab:exchange parameters}.
In that case a suitable approach is a multi-boson \sw/ theory, which describes more than four \sw/ excitations in a four-sublattice magnet. 
Out from the ortho-phosphate compounds, the multi-boson \sw/ theory has been only applied to \lcpo/, a $S=3/2$ spin system~\cite{Kocsis2018}.
Developing a multi-boson \sw/ theory for \lfpo/ is a tedious calculation, therefore, it is out of the scope of this paper.

Some of the observed features can be explained qualitatively in the limit of zero exchange and \DM/ coupling.
Assuming rotational symmetry about the $y$ axis in Eq.~(\ref{equation:model}), $ \Lambda_z=\Lambda_x$ , the spins are parallel to the quantization axis $y$, and the energy levels $E_{m_s}$  of spin $S=2$ are $E_0$, $E_{\pm1}$ and $E_{\pm2}$.
When the \hparal{y} field is applied, the energy difference $E_{+2}-E_{-2}$ increases approximately at a rate 4\,\wn/T$^{-1}$, as observed for the \sw/ excitation \mode{15}.
The electric dipole activity comes from the on-site spin-induced polarization which in the lowest order of spin operators is $P \propto \hat{S}_\alpha \hat{S}_\beta$ ($\alpha, \beta = x, y, z$)~\cite{Kocsis2018}.
Although $P\propto \hat{S}_x^2$ and $\hat{S}_z^2$ (quantization axis is $y$) couple   states different by $\Delta m_s=\pm2$ it does not explain the $|\Delta m_s|\ge 3$ transitions, \mode{10}, \mode{14} and \mode{15}.
However, in \lfpo/ the single ion anisotropies are not equal, $\Lambda_z\neq\Lambda_x$ and mix $E_0$ into $E_{\pm2}$ states, see Table\,I in Ref.\,\cite{Yiu2017}.
Therefore, the selection rule for the electric-dipole transition, $\Delta m_s=2$, and mixing of states gives finite electric-dipole moment to the $\Delta m_s=4$ transition.
In a similar manner, $P\propto \hat{S}_x\hat{S}_y$ and $\hat{S}_y\hat{S}_z$ could give rise to $\Delta m_s = \pm 1$ transitions and if the mixing of states is taken into account, then to the electric-dipole allowed $\Delta m_s = \pm 3$ transitions.

Two \sw/s,  $\omega_1(\vect{q}_1)$ and $\omega_2(\vect{q}_2)$, can be excited by THz radiation of frequency  $\omega = \omega_1 + \omega_2$  if $\vect{q}_1=-\vect{q}_2$, which is termed as two-magnon excitation.
The exact frequency dependence of this absorption depends on the coupling mechanism between the radiation and the \sw/ and on the density of \sw/ states \cite{Halley1965,Allen1966,Loudon1968,Tanaka1990,Tanabe2005,Filho2015}.
This leads to broad absorption bands with peaks at the highest density of \sw/ states \cite{Halley1965,Allen1966,Loudon1968,Fert1978,Tanaka1990,Hildebrand1999,Peedu2019}, mostly with \sw/s from the edge of the Brillouin zone.
Since the product of the two spin operators has the same time-reversal parity as the electric dipole moment, the simultaneous excitation of two spin-waves by the electric field is allowed and this mechanism usually dominates over the magnetic-dipole active absorption~\cite{Richards1967}.
A relatively broad electric-dipole active absorption line is \mode{13}. 
If $\omega_1(\vect{q}_1)=\omega_2(\vect{q}_2)$, the \sw/ frequency should  be $\omega_1 \approx 60$\wn/$=7.4$\,meV.
At about the same energy  two  dispersion curves cross in the $[0,1.5,0]$ Brillouin zone point of the two-spin unit cell~\cite{Toft-Petersen2015,Yiu2017}. 
The $[0,0.5,0]$ point, equivalent to $[0,1.5,0]$, is the Brillouin zone boundary of the four-spin unit cell and therefore we expect anti-crossing of two dispersion curves which leads to increase in the density of magnon states at this point.
Thus, considering the linewidths, energy scale, and the electric-dipole activity, \mode{13} could be a two-magnon excitation.
Another candidate for a two-magnon excitation is the electric dipole active \mode{12}.
Although it is relatively broad in zero field, it has a complicated field dependence in \hparal{y}, see Fig.\,S5 in the Supplementary Material, what can not be explained within a simple model of two-magnon excitation.

\section{Summary}

We studied the magnetic ground state and the spin excitations of the magnetoelectric antiferromagnet \lfpo/ by magnetization measurements in magnetic fields  up to 120\,T and by THz absorption spectroscopy  up to 33\,T. 
Magnetization measurements revealed a spin-flop transition at 32\,T before reaching the saturation at 56\,T.
We found 17 absorption lines below 175\wn/ (5.25\,THz) appearing in the magnetically ordered phase.
Based on the magnetic field dependence of the resonance frequencies and the intensities, we assigned four of them to magnon modes (\mode{4}-\mode{7}), eight  to multiboson \sw/ excitations (\mode{8}-\mode{11}, \mode{14}-\mode{17}), two  to two-magnon excitations (\mode{12}, \mode{13}) and the rest to the absorption by impurity spins (\mode{1}-\mode{3}).
We applied a mean-field model, which describes well the four magnon modes (\mode{4}-\mode{7}). 
We attribute the other \sw/ modes to excitations with $|\Delta m_s|>1$ arising due to the large, $S=2$, spin of octahedrally coordinated Fe$^{2+}$ ions. 
Such excitations may become electric-dipole active due to symmetry allowed coupling between spin-quadrupolar fluctuations and electric polarization.
Two modes, \mode{7} and \mode{14}, are magneto-electric resonances with  significant coupling to both, electric and magnetic field component of radiation.
Additional experiments on magneto-electrically poled samples are needed to clarify if these two resonances show non-reciprocal directional dichroism~\cite{Kocsis2018,Kocsis2019}.

    \section{Acknowledgments}
	The authors acknowledge the valuable discussions with Karlo Penc and thank Kirill Amelin and Jakub Vit for fruitful discussions and for the help with the THz spectroscopy measurements.
	This project was supported by the Estonian Research Council grant PRG736, institutional research funding IUT23-3 of the Estonian Ministry of Education and Research, the European Regional Development Fund project TK134, by the bilateral program of the Estonian
and Hungarian Academies of Sciences under Contract No. NMK2018-47, by the Hungarian National Research, Development and Innovation Office—NKFIH Grants No. FK 135003.
	The high magnetic field magnetization experiments were supported by LNCMI-CNRS and HFML-RU/NWO-I, members of the European Magnetic Field Laboratory (EMFL).
	D. S. acknowledges the FWF Austrian Science Fund I 2816-N27 and TAI 334-N.
	The cooperation of Austrian and Hungarian partners was supported by Austrian Agency for International Cooperation in Education Research Grant No. WTZ HU 08/2020 and by the Hungarian NKFIH Grant No. 2019-2.1.11-TÉT-2019-00029.
	V. K. was supported by the RIKEN Incentive Research Project and B.B. acknowledges the support by the European Research Council (Grant Agreement No.835279-Catch-22).
	The data handling, calculations and figures were done in Python programming language using libraries NumPy \cite{2020NumPy-Array}, Matplotlib \cite{Matplotlib2007}, Scipy \cite{2020SciPy-NMeth} and Pandas \cite{McKinney2010}.
	
	L.P., V.K., and D.S. contributed equally to this work.


%

\end{document}


\title{Refining magnetic interactions from the magnetic field dependence of spin-wave excitations in magnetoelectric \lfpo/}

	\author{L. Peedu}
	\affiliation{National Institute of Chemical Physics and Biophysics, Akadeemia tee 23, 12618 Tallinn, Estonia}
	
	\author{V. Kocsis}
	\affiliation{RIKEN Center for Emergent Matter Science (CEMS), Wako, Saitama 351-0198, Japan}
	
	\author{D. Szaller}
	\affiliation{Institute of Solid State Physics, TU Wien, 1040 Vienna, Austria}
	
	\author{B. Forrai}
    \affiliation{Department of Physics, Institute of Physics, Budapest University of Technology and Economics, M\H{u}egyetem rkp. 3., H-1111 Budapest, Hungary}
	
	\author{S. Bord\'acs}
    \affiliation{Department of Physics, Institute of Physics, Budapest University of Technology and Economics, M\H{u}egyetem rkp. 3., H-1111 Budapest, Hungary}
	
	\author{I. K\'ezsm\'arki}
    \affiliation{Department of Physics, Institute of Physics, Budapest University of Technology and Economics, M\H{u}egyetem rkp. 3., H-1111 Budapest, Hungary}
	\affiliation{Experimental Physics 5, Center for Electronic Correlations and Magnetism,Institute of Physics, University of Augsburg, 86159 Augsburg, Germany}
	
	\author{J. Viirok}
	\affiliation{National Institute of Chemical Physics and Biophysics, Akadeemia tee 23, 12618 Tallinn, Estonia}
	
	\author{U. Nagel}
	\affiliation{National Institute of Chemical Physics and Biophysics, Akadeemia tee 23, 12618 Tallinn, Estonia}
	
	\author{B. Bern\'ath}
    \author{D. L. Kamenskyi}
    \affiliation{High Field Magnet Laboratory (HFML-EMFL), Radboud University, Toernooiveld 7, 6525 ED Nijmegen, The Netherlands}
	
	\author{A. Miyata}
	\author{O. Portugall}

	\affiliation{Laboratoire National des Champs Magn\'etiques Intenses (LNCMI-EMFL), CNRS-UGA-UT3-INSA,143 Avenue de Rangueil, 31400 Toulouse, France}
		
	\author{Y. Tokunaga}
	\affiliation{RIKEN Center for Emergent Matter Science (CEMS), Wako, Saitama 351-0198, Japan}
	\affiliation{Department of Advanced Materials Science, University of Tokyo, Kashiwa 277-8561, Japan}
	
	\author{Y. Tokura}
	\affiliation{RIKEN Center for Emergent Matter Science (CEMS), Wako, Saitama 351-0198, Japan}
	\affiliation{Department of Applied Physics, University of Tokyo, Hongo, Tokyo 113-8656, Japan}
	
	\author{Y. Taguchi}
	\affiliation{RIKEN Center for Emergent Matter Science (CEMS), Wako, Saitama 351-0198, Japan}
	
	\author{T. R{\~o}{\~o}m}
	\affiliation{National Institute of Chemical Physics and Biophysics, Akadeemia tee 23, 12618 Tallinn, Estonia}

	\date{\today }
	\begin{abstract}
		
		
	\end{abstract}
	\maketitle

\section{THz absorption spectra of {\protect\lfpo/} in magnetic field}

Fig.\ref{fig:LFPO_H_x_all_configurations}, \ref{fig:LFPO_H_y_all_configurations} and \ref{fig:LFPO_H_z_all_configurations} show the THz absorption spectra measured for three applied magnetic field directions.
\begin{center}
	\renewcommand{\thefigure}{S1}
	\begin{figure*}			
			\includegraphics[width=15truecm]{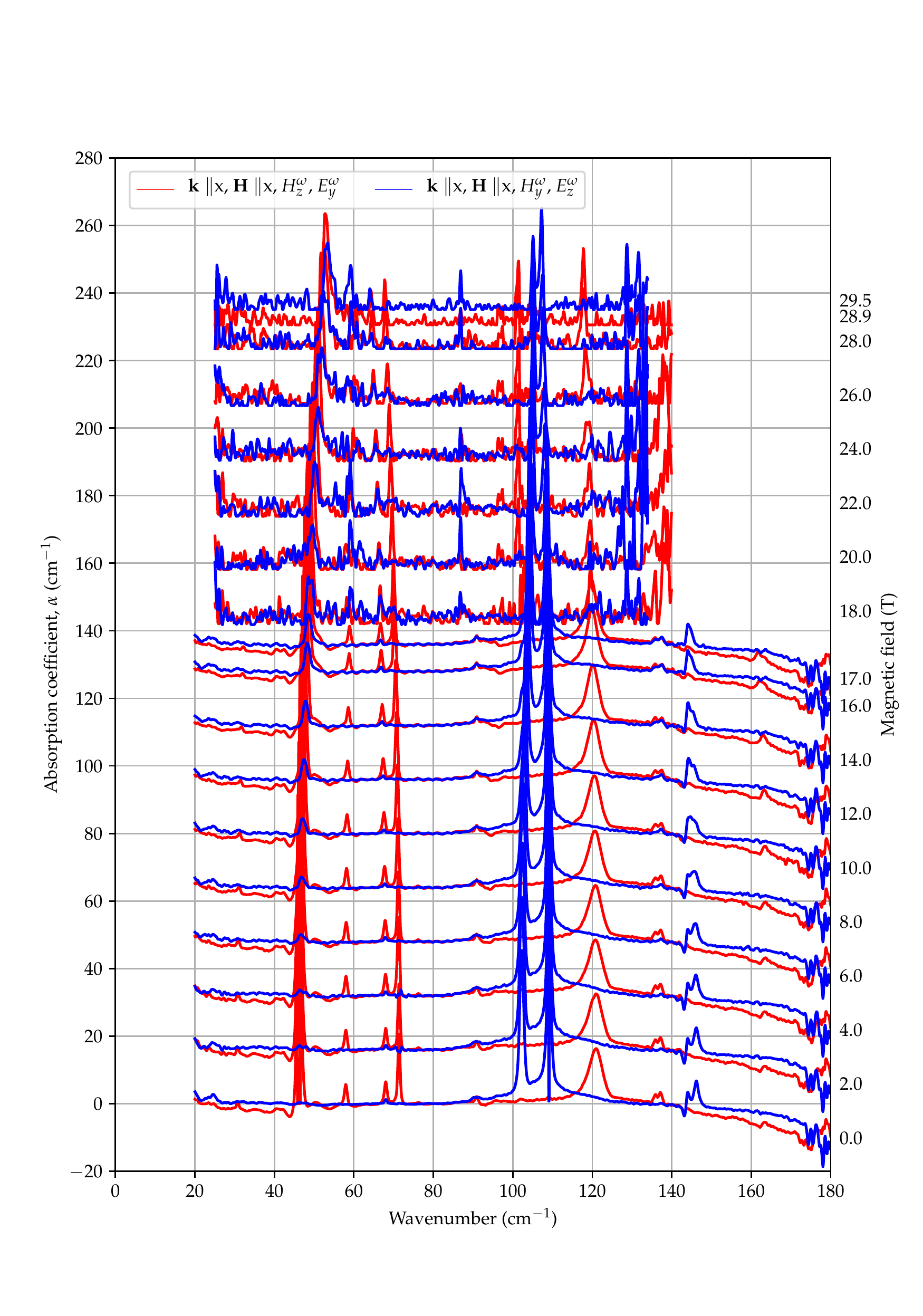}
			\caption{
					\lfpo/ THz absorption spectra of spin-wave excitations in magnetic field \hparal{x} at $T$ = 3.5\,K in two orthogonal polarizations, $\{H^\omega_y, E^\omega_z\}$ and $\{H^\omega_z, E^\omega_y\}$.
					The spectra are shifted to zero between 73 to 85\wn/ and an offset proportional to the magnitude of the magnetic field is added. 
					The spectra up to 17\,T are calculated with the reference spectra from 55\,K.
					The spectra above 17\,T are calculated with statistically found baseline and the absorption represents the changes relative to the 0\,T spectra.									 
					\label{fig:LFPO_H_x_all_configurations}}
		\end{figure*}
\end{center}
\begin{center}
	\renewcommand{\thefigure}{S2}
	\begin{figure*}			
		\includegraphics[width=15truecm]{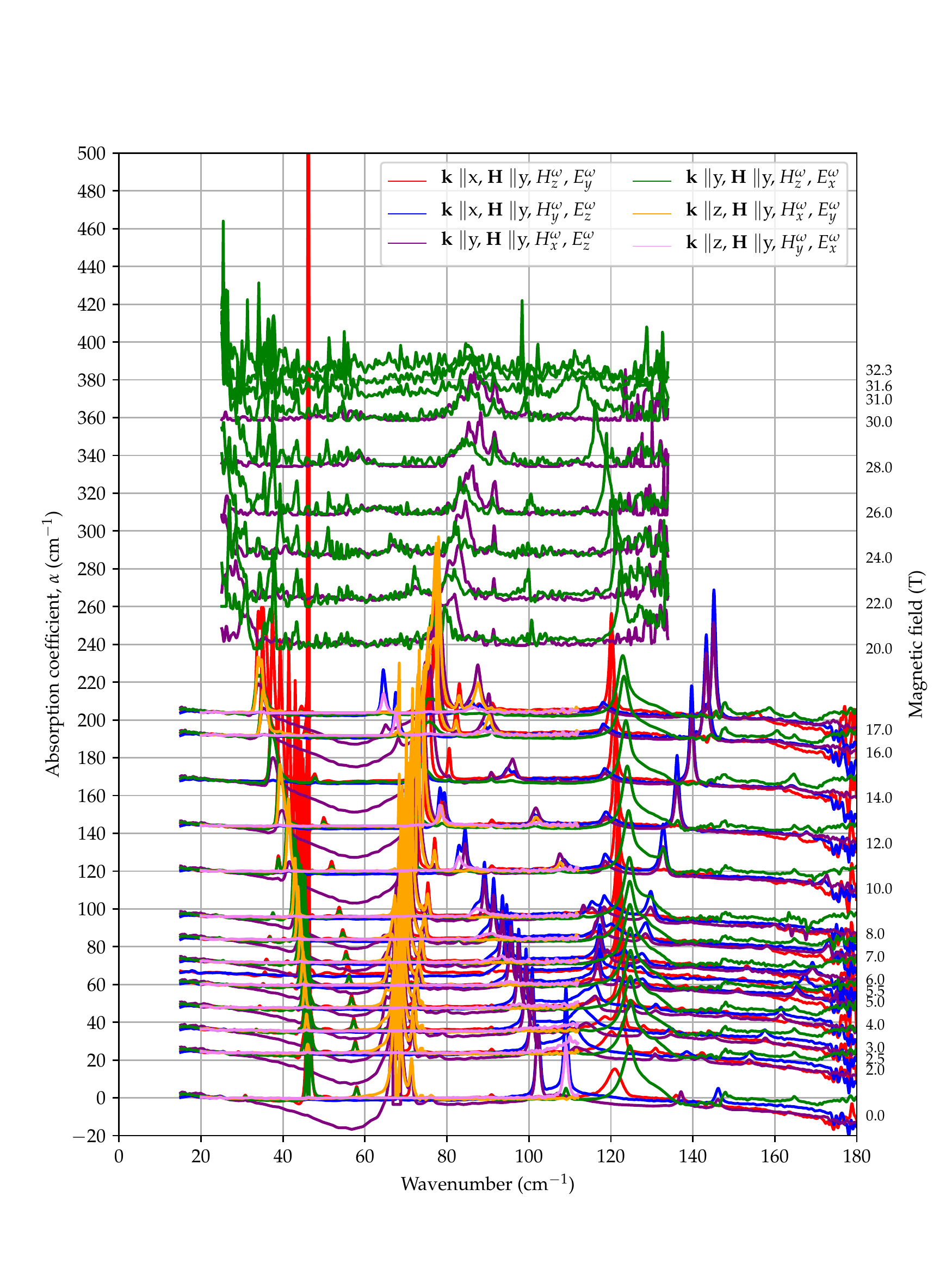}
		\caption{
			\lfpo/ THz absorption spectra of spin-wave excitations in magnetic field \hparal{y} below 4\,K.
			The colors correspond to different polarization configurations $\{H^\omega_i, E^\omega_j\}$ indicated at the top of figure.
			The spectra below 17.5\,T are shifted to zero between 20 to 30\wn/ and above 17.5\,T are shifted to zero between 60 to 65\wn/, additionally an offset proportional to the magnitude of magnetic field is added. 
			The spectra up to 17\,T are calculated with the reference spectra from 55\,K.
			The spectra above 17\,T are calculated with statistically found baseline and the absorption represents the changes relative to the 0\,T spectra.	
			\label{fig:LFPO_H_y_all_configurations}}
	\end{figure*}
\end{center}
\begin{center}
	\renewcommand{\thefigure}{S3}
	\begin{figure*}			
		\includegraphics[width=15truecm]{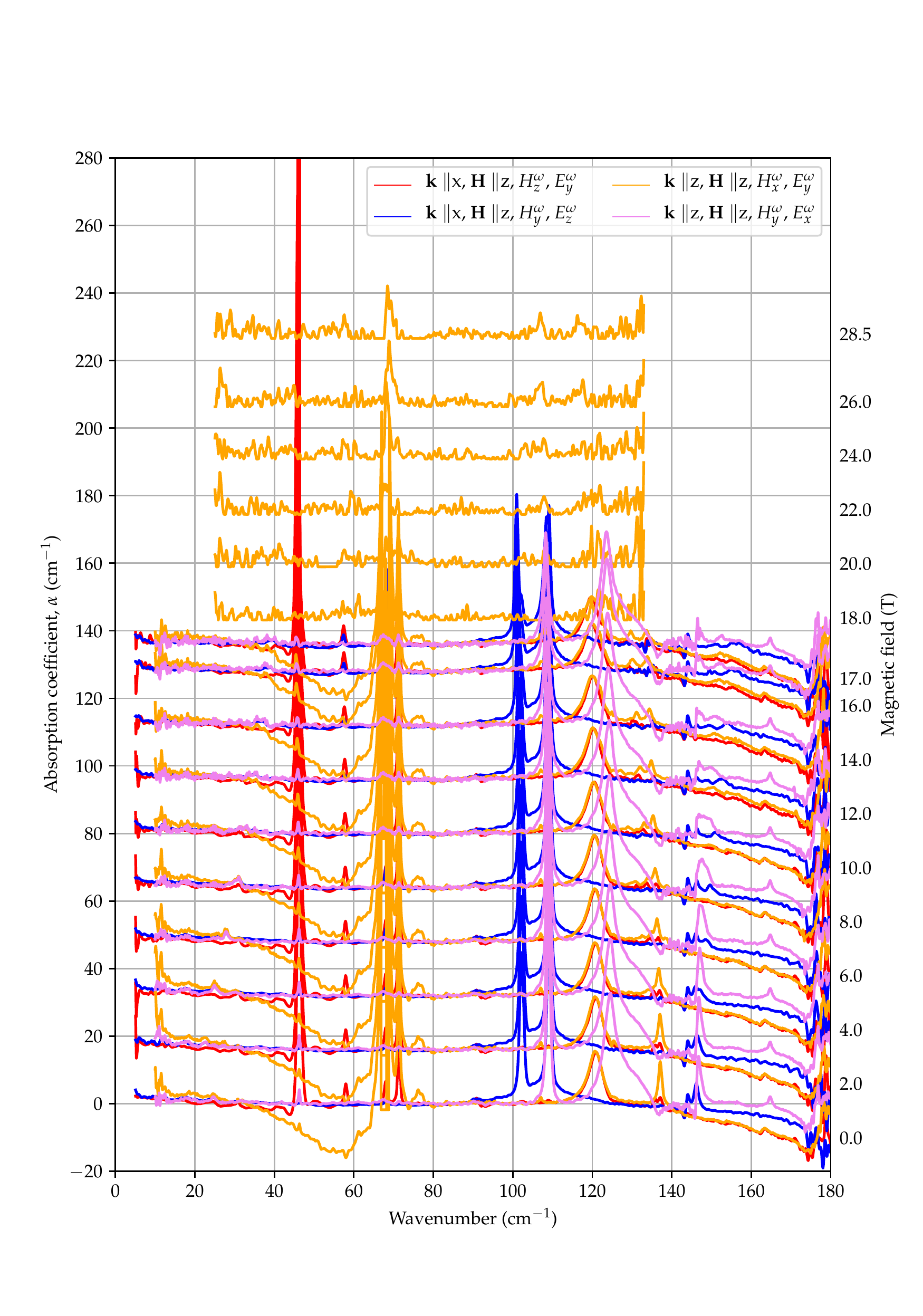}
		\caption{
			\lfpo/ THz absorption spectra of spin-wave excitations in magnetic field \hparal{z} below 4\,K.
			The colors correspond to different polarization configurations $\{H^\omega_i, E^\omega_j\}$ indicated at the top of figure.
			The spectra is shifted to zero between 80 to 90\wn/ and an offset proportional to the magnitude of magnetic field is added.
			The spectra up to 17\,T are calculated with the reference spectra from 55\,K.
			The spectra above 17\,T are calculated with statistically found baseline and the absorption represents the changes relative to the 0\,T spectra.	
			\label{fig:LFPO_H_z_all_configurations}}
	\end{figure*}
\end{center}

\section{Magnetic dipole active transitions of the mean-field model \sw/s   }

\begin{center}
	\begin{figure*}[h]	
		\renewcommand{\thefigure}{S4}
		\includegraphics[width=15truecm]{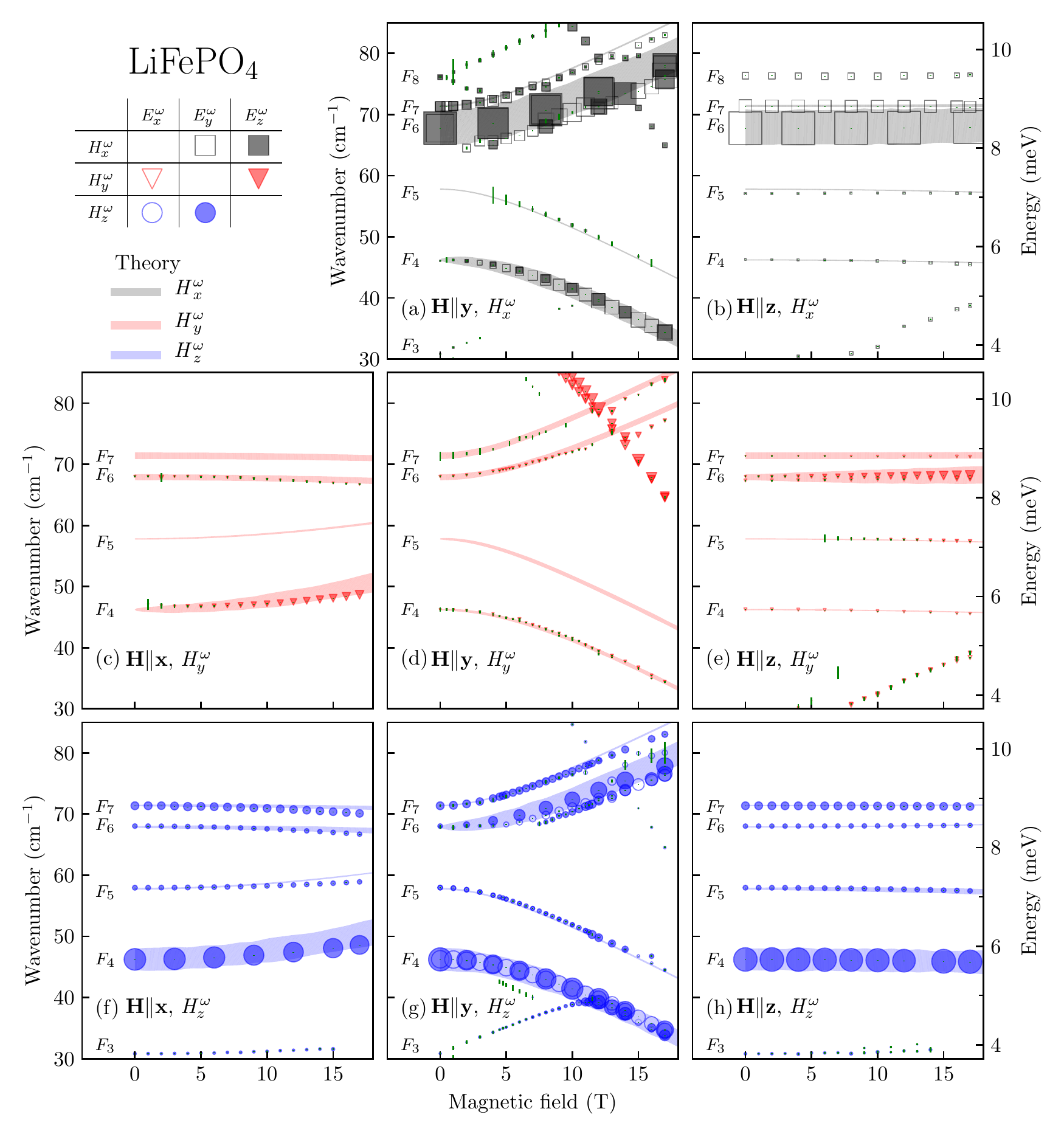} 
		\caption{
			\lfpo/ magnetic field dependence of the \sw/ resonance frequencies and absorption line areas at $T=3.5\,\mathrm{K}$. 
			The columns corresponds to individual magnetic field direction as \{\hparal{x}, \hparal{y}, \hparal{z}\} and rows correspond to the individual oscillating magnetic field direction of light as \{\hwu{x},\hwu{y},\hwu{z}\}.
			Symbols are the fit results of experimentally measured resonances and correspond to six combinations of  $\{E^\omega_i, H^\omega_j\}$ as indicated on top left of the figure. 
			The symbol height is proportional to the square root of experimental absorption line area with the same scaling as wavenumber axis.
			The symbol frequency error bar is marked with green vertical line.
			The wide horizontal solid lines are the results of the mean-field calculation, \mode{4}, \mode{5}, \mode{6} and \mode{7}, in order of increasing frequency;  the width of the line is proportional to the square root of the line area with the same scale as wavenumber axis and calculated in the magnetic dipole approximation.
			\label{fig:lfpo_9x_panels_scatter_plot}}
	\end{figure*}
\end{center}

The experimental magnetic field dependence of modes \mode{4}, \mode{5}, \mode{6} and \mode{7} frequencies and  intensities in comparison with the mean-field model results are shown in Fig.\ref{fig:lfpo_9x_panels_scatter_plot}.
Note that the data is displayed in the  limited frequency  range from  30 to 85\wn/.
Each column of figure panels corresponds to one direction of the applied magnetic field and the row corresponds to one direction of the oscillating magnetic field.
  The oscillating electric field component of experimental data can be found from the table at top left of  Fig.\,\ref{fig:lfpo_9x_panels_scatter_plot}.
The model intensities, shown as colored  lines, describe well the intensities of the strongest modes \mode{4} and \mode{6}.
For the weak absorption line  \mode{5} the theory under-estimates intensities, see Table\,\ref{tabel:F5_mode_intensity}.
The mode \mode{7} is a ME resonance and therefore there is disagreement between experimental line intensities and intensities predicted in the magnetic-dipole approximation by the mean-field model,  see Table\,\ref{tabel:F7_mode_intensity}. 

\def\arraystretch{1.5}	
\setlength{\tabcolsep}{18pt}
\begin{table}		
		\caption{Experimental $S_\mathrm{exp}$ and theoretical  $S_\mathrm{theor}$ (from the mean-field model) absorption line area of \mode{5} in different polarization configurations.
		} \label{tabel:F5_mode_intensity}	
		\vspace{5pt} 
		\begin{tabular}{ *{5}{c} }
			\hline \hline
			Sample & \ew/ & \hw/ & $S_{\mathrm{exp}}$ & $S_{\mathrm{theor}}$ \\ \hline 
			 S3B B3B & $y$ & $x$ & 1.05$\pm$0.13 & 0.002 \\ 
			 S3B B2C & $y$ & $z$ & 5.74$\pm$0.22 & 0.5 \\ 
			 S3B B2C & $y$ & $z$ & 5.6$\pm$0.4 & 0.5 \\ 
			 S3B B5 & $x$ & $z$ & 5.6$\pm$0.9 & 0.5 \\ 
			 S3B B2C & $y$ & $z$ & 5.1$\pm$0.9 & 0.5 \\ 
			 \hline \hline
		\end{tabular}
\end{table}

\begin{table}	
	
	\caption{\mode{7} intensity comparison with theory in different polarization configurations.
	} \label{tabel:F7_mode_intensity}	
	\vspace{5pt} 
	\begin{tabular}{ *{5}{c} }
		\hline \hline
		Sample & \ew/ & \hw/ & $S_{\mathrm{exp}}$ (cm$^{-2}$)& $S_{\mathrm{theor}}$ \\ \hline 
		S3B B5 & $z$ & $x$ & 20.9$\pm$0.4 & 0.8 \\ 
		S3B B3B & $y$ & $x$ & 37.09$\pm$0.27 & 0.8 \\ 
		S3B B3B & $x$ & $y$ & 0.32$\pm$0.16 & 34.1 \\ 
		S3B B2C & $z$ & $y$ & 0.06$\pm$0.28 & 34.1 \\ 
		S3B B2C & $y$ & $z$ & 14.31$\pm$0.16 & 0.001 \\ 
		S3B B2C & $y$ & $z$ & 13.9$\pm$0.3 & 0.001 \\ 
		S3B B2C & $y$ & $z$ & 14.23$\pm$0.19 & 0.001 \\ 
		S3B B5 & $x$ & $z$ & 0.31$\pm$0.17 & 0.001 \\ 
		\hline \hline
	\end{tabular}
\end{table}

\section{\mode{12} in {\protect $\mathbf{H} \parallel \mathbf{y}$}}
The mode \mode{12} has slope $-1.9$\,\wnnospace/T$^{-1}$ in \hu{y}, seen   only   in Voigt configuration, Fig.\,5(d).
Such behavior is more clearly seen in Fig.\,\ref{fig:LFPO_H[y]_BL_temp_ref_100_to_150wn_F12}. 
As this figure shows, the intensity of \mode{12} vanishes to zero above 6\,T. 
Additional \ewu{y}-active   absorption line appears on the higher frequency side  of  \mode{12} what is labeled $F_{12}^*$.
Close to \mode{12} there is  \mode{13}, but it is   \ewu{x}-active instead.
One can also see from Fig.\,\ref{fig:LFPO_H_y_all_configurations} that  $F_{12}^*$ and \mode{13} are different modes because they appear at different frequency. 
Thus there is an additional \sw/ excitation between \mode{12} and \mode{13} with zero intensity in zero magnetic field.
It becomes \ewu{y}-active above few tesla in magnetic field \hparal{y}.
 
	
\begin{center}
	\renewcommand{\thefigure}{S5}
	\begin{figure}[h]
		
		\includegraphics[width=8truecm]{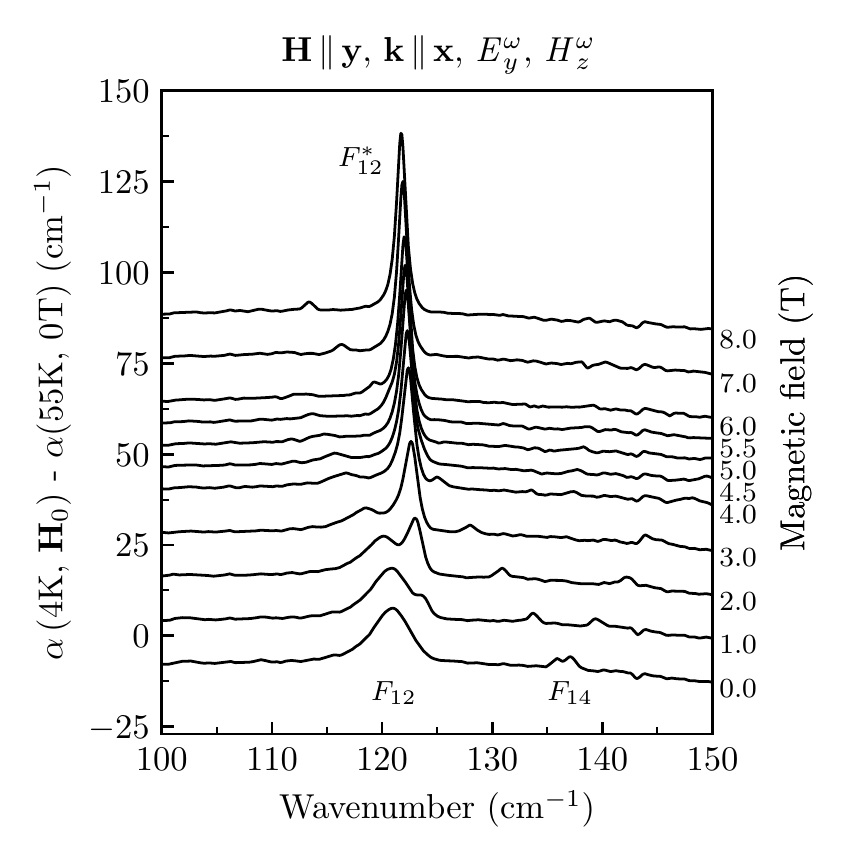}
		\caption{
			The THz spectra of \lfpo/ mode \mode{12} in magnetic field \hparal{y} at 3.5\,K.
			Polarization of THz radiation is $\{\ewu{y},\hwu{z}\}$.
			In \hparal{y} below \mode{12} there appears a narrow line $F_{12}^*$ that is almost constant in magnetic field.
			This figure helps to understand the complicated magnetic field dependence of \mode{12}.			
			\label{fig:LFPO_H[y]_BL_temp_ref_100_to_150wn_F12}}
	\end{figure}
\end{center}



\section{Exact diagonalization of a spin cluster Hamiltonian}\label{sec:QM_model}

Exact numerical diagonalization (ED) of the same Hamiltonian as in the mean-field approach,  Eq.~(1),  was performed on   a four-spin cluster.
The basis has 625 states $\ket{m_{1s},m_{2s},m_{3s},m_{4s}}$ where $m_{is}$ is the $i$-th spin projection quantum number on the $z$ axis, $S_i^z\ket{m_{is}} = m_{is}\ket{m_{is}}$ and  $m_{is}\in \{ -2, -1, 0, 1, 2 \}$. 	
The spectra of magnetic-dipole active transitions were calculated at  $T=55$\,K and are shown in Fig.\,\ref{fig:LFPO_QM_ED_55K_0T}.
The frequency and the selection rule of \fos/ is reproduced.

\begin{center}
	\renewcommand{\thefigure}{S6}
	\begin{figure}
		\includegraphics[width=8truecm]{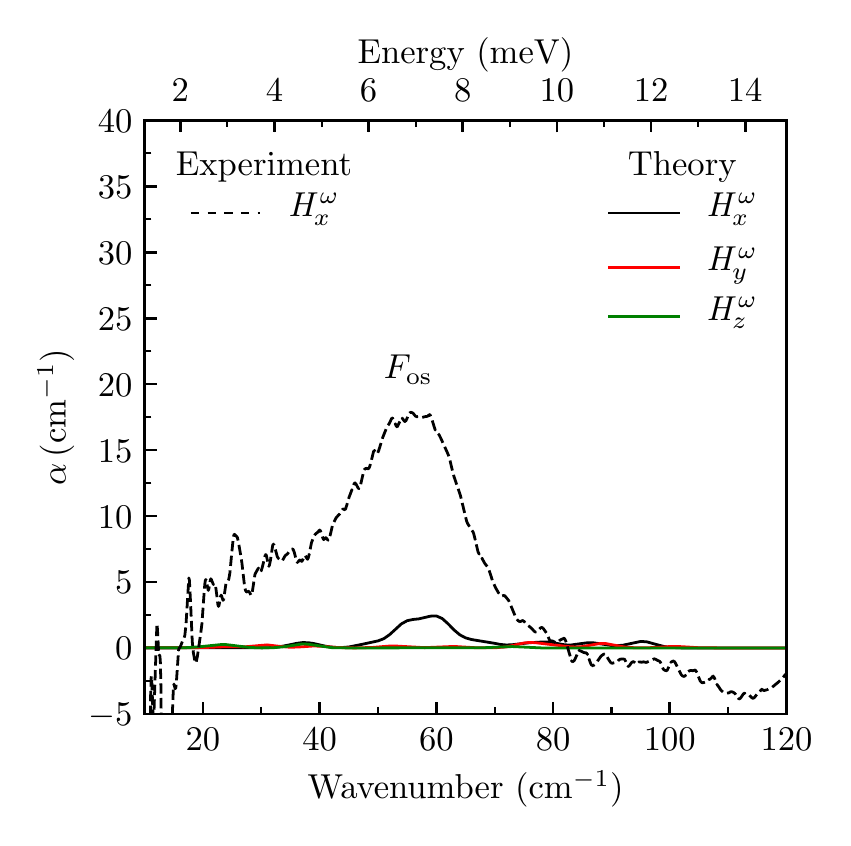}
		\caption{\label{fig:LFPO_QM_ED_55K_0T}
			THz absorption spectra from the exact diagonalization of a 4-spin cluster at 55\,K in zero applied magnetic field.
			The black, red and green spectra correspond to the transitions induced by oscillating magnetic fields components \hwu{x}, \hwu{y} and \hwu{z}, respectively.
			The spectra are the sum of individual transitions described by the  Gaussian line  shape  with  $\mathrm{FWHM}=5$\wn/.}
	\end{figure}
\end{center}


\section{Magnetization normalization}\label{sec:SM_magnetization}
The semi-destructive pulses magnetization values were normalized by DC magnetization measurement.
The normalization was done by linearly fitting the DC measurements in range 4-14\,T and pulsed field measurements in range 4-20\,T, and by multiplying the pulsed field measurement each values with the ratio of the slopes obtained from linear fits.
The 0-4\,T range was neglected from fitting as there exists small hysteresis in \hparal{y}, see Fig.\ref{fig:magnetization_DC_-4_to_4T}.
The pulsed field range for fitting was extended up to 20\,T for minimizing the effect of noise, that was detected in range 8-14\,T, see Fig.\,\ref{fig:magnetization_AC_pulsed_magnetization}.
\begin{center}
	\renewcommand{\thefigure}{S7}
	\begin{figure}
		\includegraphics[width=8truecm]{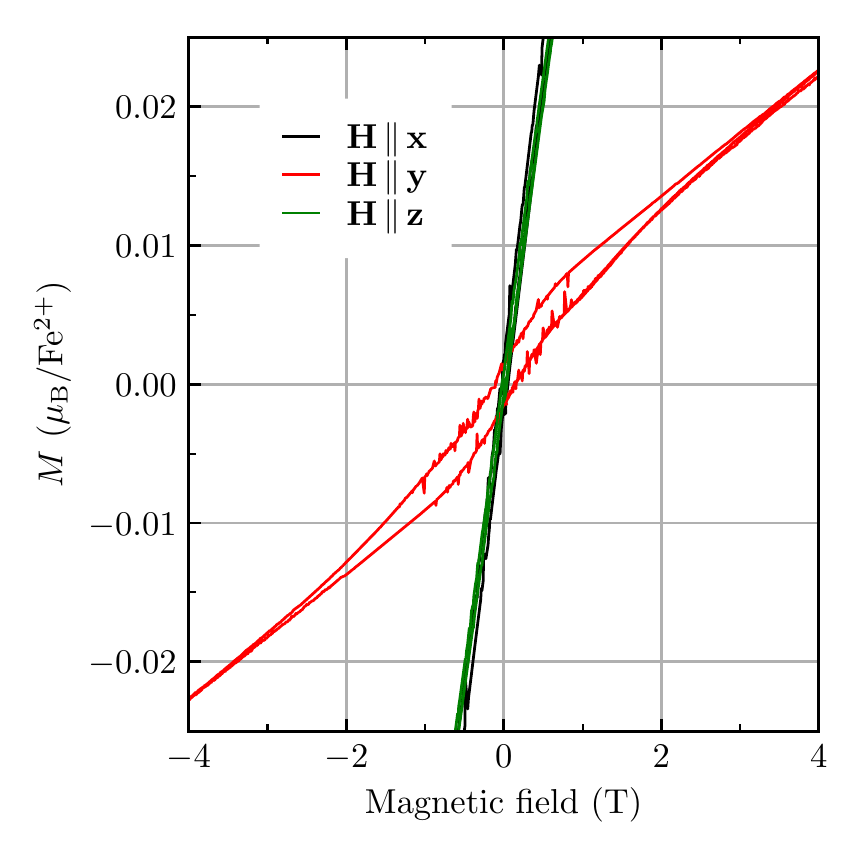}
\caption{\label{fig:magnetization_DC_-4_to_4T}
	The DC magnetic field magnetization of \lfpo/ at 3.5\,K.
	The magnetic field directions are \hparal{x} (black), \hparal{y} (red) and \hparal{z} (green).
	Along the easy axis direction, \hparal{y}, \lfpo/ shows hysteretic behaviour in low magnetic fields.}
	\end{figure}
\end{center}

\begin{center}
	\renewcommand{\thefigure}{S8}
	\begin{figure}		
		
\includegraphics[width=8truecm]{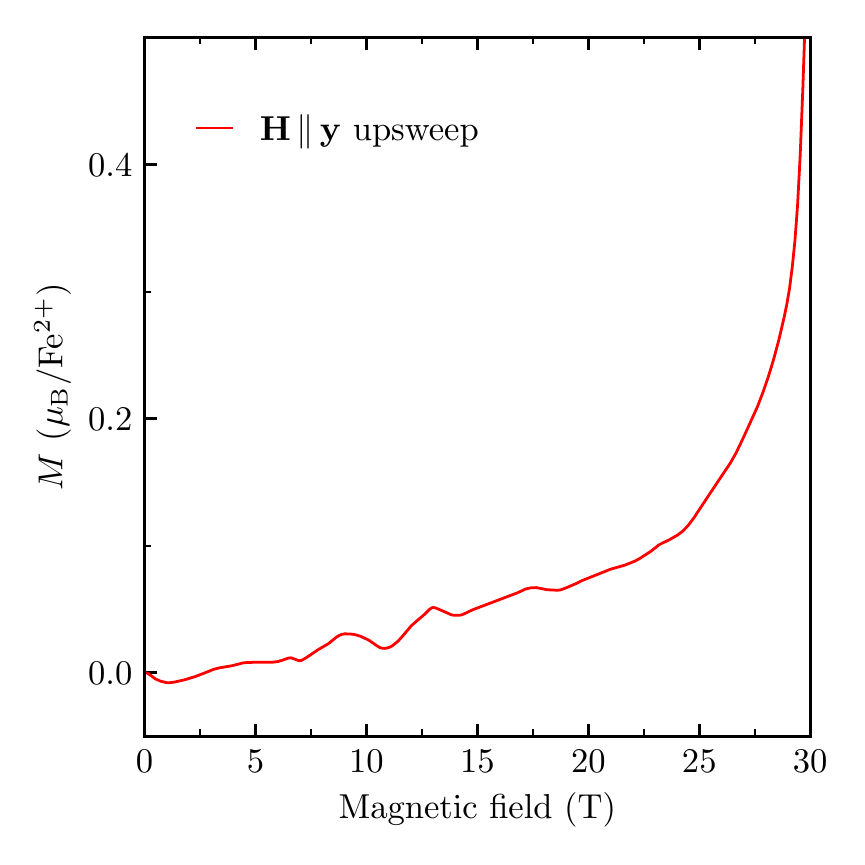}
\caption{\label{fig:magnetization_AC_pulsed_magnetization}
	The normalized semi-destructive low magnetic field upsweep magnetization of \lfpo/ at 4.2\,K.
	The magnetic field direction is \hparal{y}.
	The slopes of magnetization was normalized in range of 4 to 20\,T with the DC magnetization in range of 4 to 14\,T.
	Steep increase from linear magnetization starts at 22\,T.
	From 8 to 14\,T there exists oscillation of magnetization because of it the slope determination was extended to 20\,T.	
	} 
	\end{figure}
\end{center}


%